\documentclass[pdflatex,sn-mathphys-num]{sn-jnl}

\usepackage{graphicx}%
\usepackage{multirow}%
\usepackage{amsmath,amssymb,amsfonts}%
\usepackage{amsthm}%
\usepackage{mathrsfs}%
\usepackage[title]{appendix}%
\usepackage{xcolor}%
\usepackage{textcomp}%
\usepackage{manyfoot}%
\usepackage{booktabs}%
\usepackage{algorithm}%
\usepackage{algorithmicx}%
\usepackage{booktabs}
\usepackage{algpseudocode}%
\usepackage{listings}%
\usepackage{tabularray}
\usepackage{longtable}
\usepackage{caption}
\usepackage{float}
\usepackage{geometry}
\usepackage{tabulary,longtable}
\usepackage{booktabs, makecell, longtable}

\theoremstyle{thmstyleone}%
\theoremstyle{thmstyletwo}%
\theoremstyle{thmstylethree}%
\begin{document}

\title[Mass Spectra of Doubly Heavy Strange baryons $\Omega_{QQ^{'}}$ and their Decay properties]{Mass Spectra of Doubly Heavy Strange baryons $\Omega_{QQ^{'}}$ and their Decay properties}


\author[1]{\fnm{Akram} \sur{Ansari}}\email{akramansari78667@gmail.com}

\author*[2]{\fnm{Chandni} \sur{Menapara}}\email{chandni.menapara@gmail.com}

\author[1]{\fnm{Ajay Kumar} \sur{Rai}}\email{raiajayk@gmail.com}

\affil[1]{\orgdiv{Department of Physics}, \orgname{Sardar Vallabhbhai National Institute of Technology}, \city{Surat}, \postcode{395007}, \state{Gujarat}, \country{India}}

\affil[2]{\orgdiv{Department of Physics}, \orgname{The Maharaja Sayajirao University of Baroda}, \city{Vadodara}, \postcode{390002}, \state{Gujarat}, \country{India}}


\abstract{This study investigates baryons $\Omega_{cc}$, $\Omega_{bc}$ and $\Omega_{bb}$ composed of two heavy quarks and one strange light quark, focusing on their radial and excited state mass spectra. We employ the Hyper Central Constituent Quark Model (hCQM), incorporating higher-order corrections, including second-order mass corrections with explicit consideration of spin-dependent interactions. This approach improves the accuracy of spin splitting predictions. Additionally, we present a detailed comparison between our predicted resonance masses and those obtained from other theoretical models. We assign all possible spin-parity quantum numbers, offering a comprehensive understanding of the properties of doubly heavy baryons. Furthermore, we compute key properties such as magnetic moments, transition magnetic moments, radiative decay widths. The analysis also reveals a linear behavior in the Regge trajectories.}




\maketitle

\section{Introduction}\label{sec1}

The advent of new data from world-renowned experimental collaborations such as LHCb, BABAR, FOCUS, CMS, Belle, and SELEX \cite{LHCb:2020iko,Koshkarev:2016rci,BaBar:2006bab,LHCb:2013hvt,Belle:2013htj,Ratti:2003ez,CMS:2021rvl} have brought renewed focus to the study of baryons containing heavy quarks (charm and bottom), positioning them at the forefront of particle physics research. The characterization of doubly heavy baryons involves examining their interactions with light or strange quarks, which couple with the doubly heavy diquark to form a complete baryonic system. Investigating these particles is essential for advancing our understanding of hadron spectroscopy, as it aids in the classification and analysis of the mass spectra, internal structure, and interaction dynamics of heavy-quark baryons. Moreover, such studies provide a valuable framework for probing the fundamental aspects of Quantum Chromodynamics (QCD).\\

\noindent	Doubly heavy baryons are hadrons composed of two heavy quarks ($QQ^{'}$ = cc, bc, or bb) and one light quark (q = u, d, or s). The pair of heavy quarks can be considered as forming a compact ``heavy diquark" core. These baryons are classified into two families based on the type of light quark they contain: The non-strange family (denoted as $\Xi_{QQ^{'}}$) includes an up (u) or down (d) quark alongside the heavy diquark, resulting in baryons such as $\Xi_{cc}$, $\Xi_{bc}$, and $\Xi_{bb}$.The strange family (denoted as $\Omega_{QQ^{'}}$) contains a strange (s) quark, leading to baryons like $\Omega_{cc}$, $\Omega_{bc}$, and $\Omega_{bb}$.As components of hadronic matter, these doubly heavy baryons have been studied using various theoretical and experimental approaches, attracting considerable interest within the particle physics community.\\
\newline
The $\Xi_{cc}^{++}$ baryon remains the sole experimentally observed member of the doubly heavy baryon family \cite{ParticleDataGroup:2024cfk}. Its initial observation was made through the decay channel  $\Xi_{cc}^{++}$$\rightarrow$$\Lambda_{c}^{+}$$K^{-}$$\pi^{+}$$\pi^{-}$ \cite{LHCb:2017iph}, and its existence was subsequently confirmed via the decay mode $\Xi_{cc}^{++}$$\rightarrow$ $\Xi_{c}^{+}$$\pi^{+}$ \cite{LHCb:2018pcs}.
The ground-state baryons containing one bottom (b) quark and one charm (c) quark specifically, the $\Omega_{bc}^{0}$(bcs), $\Xi_{bc}^{0}$(bcd), and $\Xi_{bc}^{+}$(bcu) states have been extensively investigated within various theoretical frameworks. Most theoretical models predict the masses of the $\Omega_{bc}$ and $\Xi_{bc}$ baryons to lie within the range of 6700 to 7200 $\frac{MeV}{c^{2}}$ \cite{Ebert:2002ig,Zhang:2008rt,Azizi:2014jxa,Shah:2017liu,Brown:2014ena}. Furthermore, the lifetime of the $\Omega_{bc}^{0}$baryon is estimated to be approximately 0.22 $\pm$ 0.04 picoseconds.\\
\newline
Doubly heavy baryons have been the focus of intensive theoretical exploration, employing a diverse range of models and methods. These include the relativistic quark model \cite{Ebert:2002ig,Lu:2017meb} and the non-relativistic quark model \cite{Karliner:2014gca,Albertus:2006ya,Ghalenovi:2022dok}, which offer foundational perspectives on quark dynamics. More sophisticated approaches such as effective field theory \cite{Soto:2021cgk} and the hypercentral constituent quark model (hCQM) \cite{Shah:2017liu,Mohajery:2018qhz,Salehi:2019kvb} have provided deeper insights into their internal structure. Quantum Chromodynamics (QCD)-based techniques including QCD sum rules \cite{Zhang:2008rt,Wang:2010hs,Aliev:2019lvd,ShekariTousi:2024mso}, the relativistic quark-diquark framework \cite{Ebert:2011kk}, and the compact diquark model \cite{Valcarce:2008dr} have also played a crucial role. High-precision computational methods like lattice QCD \cite{Brown:2014ena,Padmanath:2015jea,Can:2013zpa,Mathur:2018epb} have further advanced the field, complemented by alternative models such as the bag model \cite{He:2004px}, the Salpeter equation approach \cite{Giannuzzi:2009gh}, and the Faddeev method \cite{Valcarce:2008dr}. Together, these varied strategies have significantly enriched our understanding of the spectrum and properties of doubly heavy baryons.\\
\newline
This study is centered on a detailed investigation of the mass spectra of doubly heavy $\Omega_{QQ^{'}}$ baryons. Employing the Hyper Central Constituent Quark Model (hCQM) with the inclusion of higher-order corrections, we aim to resolve the intricate mass hierarchy arising from hyperfine splitting. Beyond the core analysis, the paper explores a range of essential properties, including Regge trajectories, magnetic moments, and transition magnetic moments. The forthcoming sections outline the theoretical framework and unveil results that contribute to a deeper understanding of $\Omega_{QQ^{'}}$ baryons structure.

\section{Methodology}\label{sec2}

\begin{figure}[htb!]
	\centering
	\includegraphics[scale=0.5]{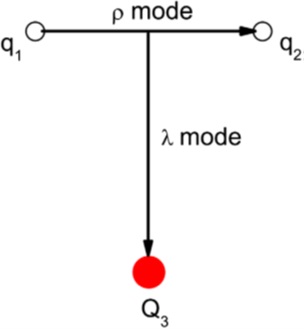}
	\caption{Diagram depicting the three inter-quark distance with two heavy and one light quark.\label{jacobi}}
\end{figure}

Doubly heavy strange baryons, composed of two heavy quarks and a strange quark, present a complex system governed by intricate three-body dynamics. Analyzing their behavior poses significant theoretical challenges due to this complexity. To make the problem more tractable, we reformulate it as an effective two-body system by introducing a parameterization that encapsulates the essential features of the three-body interaction. In this framework, we utilize Jacobian coordinates specifically, the 
$\rho$ and $\lambda$ \cite{Menapara:2023rur,Menapara:2022ksj} coordinates which allow for a more systematic treatment of the internal motion and interactions within the baryon \ref{jacobi}.
\begin{equation}
	{\rho} = \frac{1}{\sqrt{2}}(\overrightarrow{\bf r_{1}} -\overrightarrow{\bf r_{2}})\\ 
\end{equation} 
\begin{equation}
	{\lambda} = \frac{(m_{1}\overrightarrow{\bf r_{1}} + m_{2}\overrightarrow{\bf r_{2}} - (m_{1}+m_{2})\overrightarrow{\bf r_{3}})}{\sqrt{m_{1}^{2} + m_{2}^{2} + (m_{1}+m_{2})^{2}}} 
\end{equation}
Within the framework of quark dynamics, we employ a notation wherein $r_{i}$ (with $i$=1,2,3...) denotes the internal separation between quark pairs, whereas $m_{i}$ (with $i$=1,2...) signifies their individual masses. The hyperspherical coordinates, defined by the hyper radius $x$ and hyper angle $\xi$, are constructed from the Jacobi coordinates $\rho$ and $\lambda$, which are subsequently parameterized by the angular coordinates $\Omega_{\rho}$ and $\Omega_{\lambda}$. This formulation provides a mathematical basis for describing the structural properties and dynamics of quark systems.
\begin{equation}
	x = \sqrt{{{\bf \rho}^{2}} + {{\bf \lambda}^{2}}} ; \; \; \xi = arctan(\frac{\rho}{\lambda})
\end{equation}
The masses in a reduced form are provided by
\begin{equation}
	m_{\rho}=\frac{2m_1m_2}{m_1+m_2} 
	\hspace{0.5cm} and \hspace{0.5cm}
	m_{\lambda}=\frac{2m_3(m_{1}^{2}+m_{2}^{2}+m_{1}m_{2})}{(m_1+m_2)(m_1+m_2+m_3)}
\end{equation}
The operator expressing kinetic energy in the center of mass frame is formulated as \cite{Giannini:2015aya}.
\begin{equation}
	-\frac{\hbar^{2}}{2m}(\Delta_{\rho} + \Delta_{\lambda})= \frac{\hbar^{2}}{2m}(\frac{\partial^{2}}{\partial x^{2}} + \frac{5}{x}\frac{\partial}{\partial x} + \frac{L^{2}(\Omega)}{x^{2}})
\end{equation}
Here, $L^{2}(\Omega)$ refers to the quadratic Casimir operator of the six-dimensional rotational group, with eigenfunctions being hyperspherical harmonics satisfying conditions in $\Omega_{\rho}$, $\Omega_{\lambda}$, and $\xi$.
\begin{equation}
	L^{2}(\Omega_{\rho}, \Omega_{\lambda}, \xi) Y_{[\gamma]l_{\rho}l_{\lambda}}(\Omega_{\rho}, \Omega_{\lambda}, \xi) = -\gamma(\gamma + 4 )Y_{[\gamma]l_{\rho}l_{\lambda}}(\Omega_{\rho}, \Omega_{\lambda}, \xi)
\end{equation}
Since, $\gamma$ represents the grand angular quantum number, the hyperradial equation, whose solution is denoted as $\psi(x)$, is given by the following:
\begin{equation}
	\left[\frac{d^{2}}{dx^{2}} + \frac{5}{x}\frac{d}{dx} - \frac{\gamma(\gamma +4)}{x^{2}}\right]\psi(x) = -2m[E-V_{3q}(x)]\psi(x) 
\end{equation}
E serves as the eigenvalue, and $V_{3q}$ stands for the potential associated with the interaction among three quarks.\\

\noindent The model proposes the selection of a hypercentral potential, meaning it solely depends on the hyperradius x. Since the hyperradius x is influenced by all three constituent coordinates simultaneously, an hypercentral potential is not limited to a pure two-body interaction and may include three-body terms as well. Many different potentials have been employed for heavy meson and baryon systems \cite{Patel:2026xcc}. \\
The selected hypercentral potential falls into the category of hyper Coulomb type, with the addition of a linear term, denoted as $\alpha_{1} x$.
\begin{equation}
	V(x) = -\frac{\tau_{1}}{x} + \alpha_{1} x + V_{SD}(x)
\end{equation} 
Here, $\tau_{1}$ is expressed as $\frac{2}{3}\alpha_{s}$, with $\alpha_{s}$ representing the running coupling constant, while 2/3 serves as the color factor specific to baryons. Additionally, $\alpha_{1}$ corresponds to the string tension associated with baryons.\\
The spin-dependent component of Equation (8), denoted as $V_{SD}$(x), encompasses three categories of interaction terms: the spin-spin term $V_{SS}$(x), the spin-orbit term $V_{LS}$(x), and the tensor term $V_{T}$(x) \cite{Ansari:2023wlo,Ansari:2024pgk}.

\noindent The presence of the $V_{SD}$ component in the potential leads to a reduction in resonance mass with an increase in the total angular momentum quantum number J. Moreover, the inclusion of the first-order correction term $(\frac{1}{m})$ does not resolve this concern of declined mass.\\

\noindent The first-order correction term is defined as follows
\begin{equation}
	V^{1}(x)= -C_{F}C_{A}\frac{\alpha_{s}^{2}}{4x^{2}}
\end{equation}
In this context, $C_{F}$ is set at 4/3, and $C_{A}$ is set at 3, signifying the Casimir elements associated with the fundamental and adjoint representations for SU(3).\\
\begin{equation}
	\alpha_{s}= {\frac{\alpha_{s}(\mu_{0})}{1+ {(\frac{33-2 n_{f}}{12\pi})} \alpha_{s}(\mu_{0}) ln{(\frac{m_{1}+m_{2}+m_{3}}{\mu_{0}})}}}
\end{equation}

Here, $\alpha_{s}$ is 0.6 at $\mu_{0}=1$ GeV and $n_{f}=5$ is the number of active quark flavour \cite{Deur:2016tte}. $\alpha_{1}$ is string tension of the confinement part of the potential. Also, $\alpha_{1}$ is obtained by fixing the value using experimental ground state mass of the baryon. Here, the ground state reference has been taken from different models.
\noindent  
The spin-orbit masses were earlier not observed to follow the increasing trend for a particular orbital angular momentum L. So, to improvise this higher order corrections are incorporated. But, only first-order still could not account for this correction in masses. So, second order correction which is spin-orbit coupling dependent has improved the results in the correct hierarchy of masses which increase with spin-orbit term for a given state.
In this model we introduce second-order corrections to address the observed mass hierarchy attributed to spin-orbit interactions across the light to heavy sector. This correction term, characterized by its order of $O(\frac{1}{m^{2}})$, encompasses modifications for both spin-orbit and spin-tensor effects, and it is detailed as follows:
\begin{equation}
	V_{ls}^{2}(x)= (\frac{c_{s}}{2x}\frac{dV}{dx} + \frac{c_{F}}{x}(V_{1}+V_{2})){{l\cdot s}}
\end{equation}
\begin{equation}
	V_{s12}^{2}(x)=\frac{c_{F}^{2}}{3}V_{3}{{s_{12}}} 
\end{equation}
\begin{equation}
	V_{1}=-(1-\epsilon)\alpha_{1} ;\quad V_{2}=\frac{\alpha^{'}}{x^{2}}+\epsilon\alpha_{1} ;\quad  V_{3}=\frac{3\sigma_{1}}{x^{3}}
\end{equation}
The coefficients are defined as $c_{F}=1+ \frac{\alpha_{s}}{2\pi}(C_{F}+C_{A})$ and $c_{s}=2c_{F}-1$.The parameter $\alpha^{'}$ is equivalent to $\frac{2}{3}\alpha_{s}$, and $\epsilon$ and $\sigma_{1}$ represent the fitting parameters.
The expressions for $l\cdot s$ and $s_{12}$ follow a similar structure as described above. Additionally, $V^2$ is the sum of $V_{ls}^{2}(x)$ and $V_{s12}^{2}(x)$. The ultimate Hamiltonian is presented as follows:
\begin{equation}
	H = \frac{P^{2}}{2m} + V(x) + V_{SD}(x) + \frac{1}{m}V^{1}(x) + \frac{1}{m^{2}}V^{2}(x)
\end{equation}
The radial part of the Schrodinger equation with the final Hamiltonian has been solved using the Mathematica notebook.  The constituent quark masses employed in this study are outlined below:
$M_u$=338 MeV; $M_d$=350 MeV; $M_s$=500 MeV; $M_c$=1275 MeV; $M_b$=4670 MeV. The second order correction has played a role in maintaining the hierarchy of the spin-parity states of a given orbital. The uncertainty in fitting parameters has also been considered. The details are given for comparison in the results and discussion section. 

\begin{table}
	\caption{Constituent quark masses and fitting parameters}
	\begin{tabular}{ccccccc}
	\hline
	Particle & \multicolumn{3}{c}{Quark Masses(GeV)} &  Coupling Constant ($\alpha_{s}$) & String Tension ($\alpha_{1}$) &  Fitting Parameter \\
	& $m_{1}$ & $m_{2}$ & $m_{3}$ \\
	\hline
	\hline
	$\Omega_{cc}$ & 1.275 & 1.275 & 0.500 & 0.416 & 0.216 & 0.3\\
	
	$\Omega_{bc}$ & 4.67 & 1.275 & 0.500 & 0.356 & 0.216 & 0.3\\
	
	$\Omega_{bb}$ & 4.67 & 4.67 & 0.500 & 0.327 & 0.216 & 0.3 \\
	\hline
	\end{tabular}
\end{table}
\section{Results and Discussion}\label{sec3}

The family of heavy-light quark systems includes a distinct and important sector represented by doubly heavy $\Omega_{QQ^{'}}$ baryons. These baryons, which are distinguished by the presence of two heavy quarks and a light strange quark, present fascinating opportunities to investigate the dynamics of strong interactions. 
Their existence has not yet been verified by experiment and little is known about their characteristics. These particles have been the subject of extensive theoretical inquiry, with scientists using a range of models and techniques to identify their possible properties.
In this part, we investigate the mass spectra of the $\Omega_{cc}$, $\Omega_{bc}$, and $\Omega_{bb}$ baryons using the hypercentral Constituent Quark Model (hCQM) with higher-order correction techniques. The calculated mass values are presented in Tables \ref{tab1},\ref{tab2},\ref{tab3}  for $\Omega_{cc}$, $\Omega_{bc}$ and $\Omega_{bb}$  baryons, respectively.

\subsection{$\Omega_{cc}$ Baryon}
The $\Omega_{cc}$ baryon, a member of the doubly heavy baryon family, consists of a quark configuration ccs. Owing to its constituent quarks, it carries a net positive electric charge and has an isospin of I=0. In this work, we have investigated its mass spectra across a wide range of quantum states, including the 1S-5S, 1P-5P, 1D-5D, and 1F-2F states. The detailed results of our calculations are provided in Table \ref{tab1}.\\
To date, no experimental evidence has been reported for the doubly heavy $\Omega_{QQ^{'}}$ baryon. However, our theoretical predictions for the ground state are in close agreement with those obtained in earlier studies, notably matching the results in Ref. \cite{Yoshida:2015tia} and differing by approximately 30 MeV from those in Ref. \cite{Salehi:2019kvb}. Furthermore, we have analyzed the first orbital excitation with spin-parity quantum numbers $J^{P}=\frac{1}{2}^{-}$ and $J^{P}=\frac{3}{2}^{-}$, which exhibit a high degree of consistency with prior theoretical predictions, differing by 21 MeV and 29 MeV from \cite{kakadiya2023spectroscopic} and \cite{Mathur:2002ce}, respectively. Similarly, the second orbital excitation with $J^{P}=\frac{3}{2}^{+}$ and $J^{P}=\frac{5}{2}^{+}$show deviations of 44 MeV and 65 MeV from \cite{Shah:2016vmd}, respectively. The third orbital states, with $J^{P}=\frac{5}{2}^{-}$ and $J^{P}=\frac{7}{2}^{-}$, also align well with the literature, exhibiting only minor differences of 1 MeV and 53 MeV when compared with \cite{Shah:2016vmd} and \cite{Salehi:2019kvb}. These results provide valuable insights into the mass spectrum of the $\Omega_{cc}$ baryon and contribute to the growing body of theoretical work in anticipation of future experimental discoveries.\\
A recent article by Pooja et al.\cite{Jakhad:2025zos} highlights the doubly charm baryons in light of relativistic flux tube model based on quark-diquark combination with the ground state at $3718\pm16$ MeV.
The comparison of the present results as highlighted in table shows results from many different approaches. Juhi et al. \cite{Oudichhya:2021kop} has employed Regge phenomenology to predict various excited state. Also non-relativistic harmonic oscillator model \cite{Ortiz-Pacheco:2023kjn}, lattice study \cite{Mathur:2002ce} as well as recent studies with deep neural network \cite{Ghalenovi:2024scv} and QCD sum rule \cite{Tang:2011fv} has been considered. 

\begin{longtable}{ccccccccccccc}
\caption{Predicted masses for $\Omega_{cc}$ states of ours and other approaches (in MeV); Here, \cite{kakadiya2023spectroscopic}, \cite{Shah:2016vmd} are our hCQM based with different potentials; \cite{Oudichhya:2021kop} is Regge Phenomenology; \cite{Mathur:2002ce}- Lattice QCD; \cite{Ortiz-Pacheco:2023kjn} is non-relativistic harmonic oscillator model; \cite{Salehi:2019kvb}, \cite{Yoshida:2015tia}, \cite{Ghalenovi:2024scv}, \cite{Roberts:2007ni} are various constituent quark based model}
\label{tab1}\\
\toprule
State &  $J^{P}$ & $Mass_{th}$  &  \cite{kakadiya2023spectroscopic} & \cite{Shah:2016vmd} & \cite{Salehi:2019kvb} & \cite{Yoshida:2015tia} & \cite{Roberts:2007ni} & \cite{Ortiz-Pacheco:2023kjn} & \cite{Ghalenovi:2024scv} & \cite{Mathur:2002ce} & \cite{Oudichhya:2021kop} & \cite{Tang:2011fv}\\
\midrule
\endfirsthead
\caption[]{Predicted masses for $\Omega_{cc}$ states of ours and other approaches (in MeV); Here, \cite{kakadiya2023spectroscopic}, \cite{Shah:2016vmd} are our hCQM based with different potentials without and with first order corrections; \cite{Oudichhya:2021kop} is Regge Phenomenology; \cite{Mathur:2002ce}- Lattice QCD; \cite{Ortiz-Pacheco:2023kjn} is non-relativistic harmonic oscillator model; \cite{Salehi:2019kvb}, \cite{Yoshida:2015tia}, \cite{Ghalenovi:2024scv}, \cite{Roberts:2007ni} are various constituent quark based model (continued)}\\
\toprule
State &  $J^{P}$ & $Mass_{th}$  &  \cite{kakadiya2023spectroscopic} & \cite{Shah:2016vmd} & \cite{Salehi:2019kvb} & \cite{Yoshida:2015tia} & \cite{Roberts:2007ni} & \cite{Ortiz-Pacheco:2023kjn} & \cite{Ghalenovi:2024scv} & \cite{Mathur:2002ce} & \cite{Oudichhya:2021kop} & \cite{Tang:2011fv}\\
\midrule
\endhead
\midrule
$1^{2}S_{\frac{1}{2}}$ & $\frac{1}{2}^{+}$  & 3692 & 3736 & 3650 & 3662 & 3697 & 3648 & 3766 & 3750 & 3698 & 3752 & 3630\\
$1^{2}S_{\frac{1}{2}}$ & $\frac{1}{2}^{+}$ & 3692 & 3736 & 3650 & 3662 & 3697 & 3648 & 3766 & 3750 & 3698 & 3752 &\\
$1^{4}S_{\frac{3}{2}}$ & $\frac{3}{2}^{+}$ & 3780 & 3837 & 3810 & 3677 & 3787 & 3770 & 3833 & & & 3816 & 3710\\
\midrule
	$2^{2}S_{\frac{1}{2}}$ & $\frac{1}{2}^{+}$ & 4094 & 4078 & 4028 & 3964 & 4112 & 4268 & & & & &\\
	$2^{4}S_{\frac{3}{2}}$ & $\frac{3}{2}^{+}$ & 4148 & 4110 & 4085 & 3979 & 4160 & 4334 &&&&&\\
\midrule
	$3^{2}S_{\frac{1}{2}}$ & $\frac{1}{2}^{+}$ & 4427 & 4320 & 4317 & 4177 & & & & &  && \\
	$3^{4}S_{\frac{3}{2}}$ & $\frac{3}{2}^{+}$ & 4464 & 4334 & 4345 & 4193 & & & & & & &\\
\midrule
	$4^{2}S_{\frac{1}{2}}$ & $\frac{1}{2}^{+}$ & 4730 & 4514 & 4570 & 4452
	& & & & &  &  &\\
	$4^{4}S_{\frac{3}{2}}$ & $\frac{3}{2}^{+}$ & 4758 & 4521 & 4586 & 4467
	& & & & &  & & \\
\midrule
	$5^{2}S_{\frac{1}{2}}$ & $\frac{1}{2}^{+}$ & 5012 & 4676 & 4801 & 4787 & & & & &  & &\\
	$5^{4}S_{\frac{3}{2}}$ & $\frac{3}{2}^{+}$ & 5035 & 4680 & 4811 & 4802 & & & & &  &  &\\
\midrule
\midrule
	$1^{2}P_{\frac{1}{2}}$ & $\frac{1}{2}^{-}$ & 3989 & 4011 & 3964 & 3965 & 4009 & & 3970 & 4106 \footnote[1]{$J^{P}$ value is not assigned in the reference, but belongs to P-wave as mentioned.} & & 3975 &\\
    $1^{2}P_{\frac{3}{2}}$ & $\frac{3}{2}^{-}$ & 4004 & 4004 & 3948 & 3966 & & &  4002 & & & &\\
    $1^{4}P_{\frac{1}{2}}$ & $\frac{1}{2}^{-}$ & 3982 & 4014 & 3972 & 3987 & & & 4145 & & & & \\
    $1^{4}P_{\frac{3}{2}}$ & $\frac{3}{2}^{-}$ & 4001 & 4007 & 3981 & 3978 & & & 4177 & & & &\\
    $1^{4}P_{\frac{5}{2}}$ & $\frac{5}{2}^{-}$ & 4014 & 3998 & 3935 & 3926 & & & 4232 & & & 4094 &\\
    \midrule
    $2^{2}P_{\frac{1}{2}}$ & $\frac{1}{2}^{-}$ & 4321 & 4253 & 4241 & 4194 & 4101 &&&&&& \\
    $2^{2}P_{\frac{3}{2}}$ & $\frac{3}{2}^{-}$ & 4333 & 4248 & 4228 & 4155 &&&&& && \\
    $2^{4}P_{\frac{1}{2}}$ & $\frac{1}{2}^{-}$ & 4316 & 4256 & 4248 & 4215 &&&&&& &\\
    $2^{4}P_{\frac{3}{2}}$ & $\frac{3}{2}^{-}$ & 4330 & 4251 & 4234 & 4176&&&&&& &\\
    $2^{4}P_{\frac{5}{2}}$ & $\frac{5}{2}^{-}$ & 4342 & 4244 & 4216 & 4156 &&&&&& &\\
    \midrule
    $3^{2}P_{\frac{1}{2}}$ & $\frac{1}{2}^{-}$ &  4626 & 4453 & 4492 & 4453&&&&&& &\\
    $3^{2}P_{\frac{3}{2}}$ & $\frac{3}{2}^{-}$ &  4636 & 4450 & 4479 & 4444&&&&&& &\\
    $3^{4}P_{\frac{1}{2}}$ & $\frac{1}{2}^{-}$ &  4621  & 4454 & 4498 & 4475&&&&&& &\\
    $3^{4}P_{\frac{3}{2}}$ & $\frac{3}{2}^{-}$ &   4633 & 4451 & 4486 & 4466 &&&&&& &\\
    $3^{4}P_{\frac{5}{2}}$ & $\frac{5}{2}^{-}$ &  4643 & 4447 & 4469 & 4414 &&&&&& &\\
   \midrule
    $4^{2}P_{\frac{1}{2}}$ & $\frac{1}{2}^{-}$ & 4910 &&&&&& &&&& \\
    $4^{2}P_{\frac{3}{2}}$ & $\frac{3}{2}^{-}$ & 4919 &&&&&& &&&& \\
    $4^{4}P_{\frac{1}{2}}$ & $\frac{1}{2}^{-}$ & 4905 &&&&&& &&&&\\
    $4^{4}P_{\frac{3}{2}}$ & $\frac{3}{2}^{-}$ &  4916 &&&&&& &&&& \\
    $4^{4}P_{\frac{5}{2}}$ & $\frac{5}{2}^{-}$ &  4926 &&&&&& &&&& \\
   \midrule
    $5^{2}P_{\frac{1}{2}}$ & $\frac{1}{2}^{-}$ & 5179 &&&&&& &&&&\\
    $5^{2}P_{\frac{3}{2}}$ & $\frac{3}{2}^{-}$ & 5188 &&&&&& &&&&\\ 
    $5^{4}P_{\frac{1}{2}}$ & $\frac{1}{2}^{-}$ &  5175 &&&&&& &&&&\\
    $5^{4}P_{\frac{3}{2}}$ & $\frac{3}{2}^{-}$ & 5185 &&&&&& &&&&\\
    $5^{4}P_{\frac{5}{2}}$ & $\frac{5}{2}^{-}$ &  5194 &&&&&& &&&&\\
    \midrule
    \midrule
         $1^{2}D_{\frac{3}{2}}$ & $\frac{3}{2}^{+}$ &  4199 & 4091 & 4133 & 4156 & & & & 4414\footnote[1]{$J^{P}$ value is not assigned in the reference, but belongs to D-wave as mentioned.} & &  &\\
    	$1^{2}D_{\frac{5}{2}}$ & $\frac{5}{2}^{+}$ &  4213 & 4082 & 4113 & 4156 & & & & & & 4186 &\\
    	$1^{4}D_{\frac{1}{2}}$ & $\frac{1}{2}^{+}$ &  4181 & 4101 & 4156 & 4215 &&& & & & &\\
    	$1^{4}D_{\frac{3}{2}}$ & $\frac{3}{2}^{+}$ &  4194 & 4094 & 4141 & 4193&&&&&& & \\
    	$1^{4}D_{\frac{5}{2}}$ & $\frac{5}{2}^{+}$ &  4209 & 4086 & 4121 & 4155 &&&&&& & \\
    	$1^{4}D_{\frac{7}{2}}$ & $\frac{7}{2}^{+}$ &  4224 & 4075 & 4095 & 4086 & & & & & & 4354 & \\
    	\midrule
    	$2^{2}D_{\frac{3}{2}}$ & $\frac{3}{2}^{+}$ &  4509 & 4318 & 4389 & 4429 &&&&&& & \\
    	$2^{2}D_{\frac{5}{2}}$ & $\frac{5}{2}^{+}$ &  4522 & 4312 & 4372 & 4391 &&&&&&& \\
    	$2^{4}D_{\frac{1}{2}}$ & $\frac{1}{2}^{+}$ &  4493 & 4324 & 4407 & 4490 &&&&&&& \\
    	$2^{4}D_{\frac{3}{2}}$ & $\frac{3}{2}^{+}$ &  4505 & 4320 & 4395 & 4468 &&&&&&& \\
    	$2^{4}D_{\frac{5}{2}}$ & $\frac{5}{2}^{+}$ &  4518 & 4314 & 4378 & 4430 &&&&&&&\\
    	$2^{4}D_{\frac{7}{2}}$ & $\frac{7}{2}^{+}$ &  4532 & 4307 & 4358 & 4360 &&&&&&&\\
    	\hline
    	$3^{2}D_{\frac{3}{2}}$ & $\frac{3}{2}^{+}$ & 4800 &&&&&&&&&&\\
    	$3^{2}D_{\frac{5}{2}}$ & $\frac{5}{2}^{+}$ & 4810 &&&&&&&&&&\\
    	$3^{4}D_{\frac{1}{2}}$ & $\frac{1}{2}^{+}$ & 4787 &&&&&&&&&&\\
    	$3^{4}D_{\frac{3}{2}}$ & $\frac{3}{2}^{+}$ & 4796 &&&&&&&&&&\\
    	$3^{4}D_{\frac{5}{2}}$ & $\frac{5}{2}^{+}$ & 4807 &&&&&&&&&&\\
    	$3^{4}D_{\frac{7}{2}}$ & $\frac{7}{2}^{+}$ & 4819 &&&&&&&&&& \\
    	\hline
    	$4^{4}D_{\frac{3}{2}}$ & $\frac{3}{2}^{+}$ & 5072&&&&&&&&&& \\
    	$4^{2}D_{\frac{5}{2}}$ & $\frac{5}{2}^{+}$ & 5082 &&&&&&&&&&\\
    	$4^{4}D_{\frac{1}{2}}$ & $\frac{1}{2}^{+}$ & 5059 &&&&&&&&&& \\
    	$4^{4}D_{\frac{3}{2}}$ & $\frac{3}{2}^{+}$ & 5068 &&&&&&&&&& \\
    	$4^{4}D_{\frac{5}{2}}$ & $\frac{5}{2}^{+}$ & 5079 &&&&&&&&&&\\
    	$4^{4}D_{\frac{7}{2}}$ & $\frac{7}{2}^{+}$ & 5091 &&&&&&&&&& \\
    	\hline
    	$5^{2}D_{\frac{3}{2}}$ & $\frac{3}{2}^{+}$ & 5333 &&&&&&&&&&\\
    	$5^{2}D_{\frac{5}{2}}$ & $\frac{5}{2}^{+}$ & 5343 &&&&&&&&&& \\
    	$5^{4}D_{\frac{1}{2}}$ & $\frac{1}{2}^{+}$ & 5321 &&&&&&&&&& \\
    	$5^{4}D_{\frac{3}{2}}$ & $\frac{3}{2}^{+}$ & 5329 &&&&&&&&&& \\
    	$5^{4}D_{\frac{5}{2}}$ & $\frac{5}{2}^{+}$ & 5339 &&&&&&&&&& \\
    	$5^{4}D_{\frac{7}{2}}$ & $\frac{7}{2}^{+}$ & 5351 &&&&&&&&&& \\
    	\midrule
    		$1^{2}F_{\frac{5}{2}}$ & $\frac{5}{2}^{-}$ & 4390 & 4157 & 4287 & 4391 &&&&&& & \\
    		$1^{2}F_{\frac{7}{2}}$ & $\frac{7}{2}^{-}$ & 4406 & 4147 & 4264 & 4353 & & & & & & 4387&\\
    		$1^{4}F_{\frac{3}{2}}$ & $\frac{3}{2}^{-}$ & 4371 & 4167 & 4313 & 4467 &&&&&& &\\
    		$1^{4}F_{\frac{5}{2}}$ & $\frac{5}{2}^{-}$ & 4386 & 4160 & 4294 & 4429 &&&&&& &\\
    		$1^{4}F_{\frac{7}{2}}$ & $\frac{7}{2}^{-}$ & 4403 & 4150 & 4271 & 4360 &&&&&& &\\
    		$1^{4}F_{\frac{9}{2}}$ & $\frac{9}{2}^{-}$ & 4421 & 4139 & 4244 & 4292 & & & & & & 4599 &\\
    		\hline
    		$2^{2}F_{\frac{5}{2}}$ & $\frac{5}{2}^{-}$ & 4685 &&&&&& &&&&\\
    		$2^{2}F_{\frac{7}{2}}$ & $\frac{7}{2}^{-}$ & 4700 &&&&&&&&& &\\
    		$2^{4}F_{\frac{3}{2}}$ & $\frac{3}{2}^{-}$ & 4668 &&&&&&&&& &\\
    		$2^{4}F_{\frac{5}{2}}$ & $\frac{5}{2}^{-}$ & 4681 &&&&&&&&& &\\
    		$2^{4}F_{\frac{7}{2}}$ & $\frac{7}{2}^{-}$ & 4696 &&&&&&&&&& \\
    		$2^{4}F_{\frac{9}{2}}$ & $\frac{9}{2}^{-}$ & 4729 &&&&&&&&& &\\
    		\midrule
    		\bottomrule
\end{longtable}

\newpage
\subsection{$\Omega_{bc}$ Baryon}

\begin{longtable}{ccccccccccc}
	\caption{Predicted masses for $\Omega_{bc}$ states of ours and other approaches (in MeV); Here, \cite{kakadiya2023spectroscopic}, \cite{Shah:2016vmd} are our hCQM based with different potentials; \cite{Mathur:2002ce}- Lattice QCD; \cite{Salehi:2019kvb}, \cite{Yoshida:2015tia}, \cite{Ghalenovi:2024scv}, \cite{Roberts:2007ni} are various constituent quark based model}
	\label{tab2}\\
	\toprule
	State &  $J^{P}$ & $Mass_{theo.}$ & \cite{kakadiya2023spectroscopic} & \cite{Shah:2016vmd} & \cite{Salehi:2019kvb} & \cite{Yoshida:2015tia} & \cite{Roberts:2007ni} & \cite{Ghalenovi:2024scv} & \cite{Mathur:2002ce} & \cite{Tang:2011fv}\\
	\midrule
	\endfirsthead
	\caption[]{Predicted masses for $\Omega_{bc}$ states of ours and other approaches (in MeV); Here, \cite{kakadiya2023spectroscopic}, \cite{Shah:2016vmd} are our hCQM based with different potentials; \cite{Mathur:2002ce}- Lattice QCD; \cite{Salehi:2019kvb}, \cite{Yoshida:2015tia}, \cite{Ghalenovi:2024scv}, \cite{Roberts:2007ni} are various constituent quark based model (continued)}\\
	\toprule
	State &  $J^{P}$ & $Mass_{theo.}$ & \cite{kakadiya2023spectroscopic} & \cite{Shah:2016vmd} & \cite{Salehi:2019kvb} & \cite{Yoshida:2015tia} & \cite{Roberts:2007ni} & \cite{Ghalenovi:2024scv} & \cite{Mathur:2002ce}& \cite{Tang:2011fv} \\
	\midrule
	\endhead
		$1^{2}S_{\frac{1}{2}}$ & $\frac{1}{2}^{+}$ & 7077 & 7079 & 7136 & 7329 & 7136 & 7088 & 6900 & 6954 & 6750\\
		$1^{4}S_{\frac{3}{2}}$ & $\frac{3}{2}^{+}$ & 7144 & 7182 & 7187 & 7339 & 7187 & 7130 && & 6770\\
		\midrule
		$2^{2}S_{\frac{1}{2}}$ & $\frac{1}{2}^{+}$ & 7461 & 7435 & 7473 & 7475 && && \\
		$2^{4}S_{\frac{3}{2}}$ & $\frac{3}{2}^{+}$ & 7502 & 7469 & 7490 & 7485 &&&& \\
	\midrule
		$3^{2}S_{\frac{1}{2}}$ & $\frac{1}{2}^{+}$ & 7781 & 7688 & 7753 & 7609 &&&& \\ 
		$3^{4}S_{\frac{3}{2}}$ & $\frac{3}{2}^{+}$ & 7809 & 7703 & 7761 & 7619 &&&& \\
		\midrule
		$4^{2}S_{\frac{1}{2}}$ & $\frac{1}{2}^{+}$ & 8073 & 7895 & 8004 & 7782 &&&&  \\
		$4^{4}S_{\frac{3}{2}}$ & $\frac{3}{2}^{+}$ & 8094 & 7903 & 8009 & 7792 &&&& \\
		\midrule
		$5^{2}S_{\frac{1}{2}}$ & $\frac{1}{2}^{+}$ & 8346  & 8073 & 8236 & 7993 &&&& \\
		$5^{4}S_{\frac{3}{2}}$ & $\frac{3}{2}^{+}$ & 8363 & 8077 & 8239 & 8002  &&&& \\
		\midrule
		\midrule
			$1^{2}P_{\frac{1}{2}}$ & $\frac{1}{2}^{-}$ & 7357  & 7369 & 7375 & 7476 & & & 7445 &\\
			$1^{2}P_{\frac{3}{2}}$ & $\frac{3}{2}^{-}$ & 7369  & 7364 & 7363 & 7470 &&&& \\
			$1^{4}P_{\frac{1}{2}}$ & $\frac{1}{2}^{-}$ & 7331 & 7371 & 7381 & 7490  &&&& \\
			$1^{4}P_{\frac{3}{2}}$ & $\frac{3}{2}^{-}$ & 7347  & 7366 & 7369 & 7486 &&&& \\
			$1^{4}P_{\frac{5}{2}}$ & $\frac{5}{2}^{-}$ & 7359  & 7360 & 7353 & 7451&&&&  \\
			\midrule
			$2^{2}P_{\frac{1}{2}}$ & $\frac{1}{2}^{-}$ & 7610  & 7620 & 7657 & 7619&&&& \\
			$2^{2}P_{\frac{3}{2}}$ & $\frac{3}{2}^{-}$ & 7619  & 7616 & 7647 & 7595&&&& \\
			$2^{4}P_{\frac{1}{2}}$ & $\frac{1}{2}^{-}$ & 7607  & 7621 & 7662 & 7633 &&&& \\
			$2^{4}P_{\frac{3}{2}}$ & $\frac{3}{2}^{-}$ & 7617  & 7618 & 7652 & 7610 &&&& \\
			$2^{4}P_{\frac{5}{2}}$ & $\frac{5}{2}^{-}$ & 7625  & 7614 & 7639 & 7594 &&&& \\
			\midrule
			$3^{2}P_{\frac{1}{2}}$ & $\frac{1}{2}^{-}$ &  7890 & 7832 & 7912 & 7782 &&&& \\
			$3^{2}P_{\frac{3}{2}}$ & $\frac{3}{2}^{-}$ &  7897 & 7830 & 7903 & 7777 &&&& \\
			$3^{4}P_{\frac{1}{2}}$ & $\frac{1}{2}^{-}$ &  7888 & 7833 & 7916 & 7796 &&&& \\
			$3^{4}P_{\frac{3}{2}}$ & $\frac{3}{2}^{-}$ &   7895 & 7831 & 7908 & 7793 &&&& \\
			$3^{4}P_{\frac{5}{2}}$ & $\frac{5}{2}^{-}$ &  7901  & 7828 & 7896 & 7758 &&&& \\
			\midrule
			\\
			\\
			$4^{2}P_{\frac{1}{2}}$ & $\frac{1}{2}^{-}$ & 8151  &&&&&&&\\
			$4^{2}P_{\frac{3}{2}}$ & $\frac{3}{2}^{-}$ & 8157 &&&&&&&\\
			$4^{4}P_{\frac{1}{2}}$ & $\frac{1}{2}^{-}$ & 8148 &&&&&&&\\
			$4^{4}P_{\frac{3}{2}}$ & $\frac{3}{2}^{-}$ &  8156 &&&&&&&\\
			$4^{4}P_{\frac{5}{2}}$ & $\frac{5}{2}^{-}$ &  8162 &&&&&&&\\
			\midrule
			$5^{2}P_{\frac{1}{2}}$ & $\frac{1}{2}^{-}$ & 8399 &&&&&&&\\
			$5^{2}P_{\frac{3}{2}}$ & $\frac{3}{2}^{-}$ & 8405 &&&&&&&\\ 
			$5^{4}P_{\frac{1}{2}}$ & $\frac{1}{2}^{-}$ &  8396 &&&&&&&\\
			$5^{4}P_{\frac{3}{2}}$ & $\frac{3}{2}^{-}$ & 8403 &&&&&&&\\
			$5^{4}P_{\frac{5}{2}}$ & $\frac{5}{2}^{-}$ &  8409 &&&&&&&\\
			\midrule
			\midrule
				$1^{2}D_{\frac{3}{2}}$ & $\frac{3}{2}^{+}$ &  7500 & 7695 & 7545 & 7597 & & & 7754& \\
				$1^{2}D_{\frac{5}{2}}$ & $\frac{5}{2}^{+}$ &  7508 & 7452 & 7531 & 7571 &&&&\\
				$1^{4}D_{\frac{1}{2}}$ & $\frac{1}{2}^{+}$ &  7489 & 7464 & 7562 & 7634 &&&&\\
				$1^{4}D_{\frac{3}{2}}$ & $\frac{3}{2}^{+}$ &  7497 & 7454 & 7536 & 7618 &&&&\\
				$1^{4}D_{\frac{5}{2}}$ & $\frac{5}{2}^{+}$ &  7506 & 7454 & 7536 & 7595 &&&&\\
				$1^{4}D_{\frac{7}{2}}$ & $\frac{7}{2}^{+}$ &  7515 & 7447 & 7518 & 7552 &&&&\\
				\hline
				$2^{2}D_{\frac{3}{2}}$ & $\frac{3}{2}^{+}$ &  7784 & 7695 & 7807 & 7768 &&&&\\
				$2^{2}D_{\frac{5}{2}}$ & $\frac{5}{2}^{+}$ &  7792 & 7691 & 7795 & 7744 &&&&\\
				$2^{4}D_{\frac{1}{2}}$ & $\frac{1}{2}^{+}$ &  7773 & 7700 & 7821 & 7806 &&&&\\
				$2^{4}D_{\frac{3}{2}}$ & $\frac{3}{2}^{+}$ &  7781 & 7697 & 7812 & 7792 &&&&\\
				$2^{4}D_{\frac{5}{2}}$ & $\frac{5}{2}^{+}$ &  7790 & 7692 & 7799 & 7767 &&&&\\
				$2^{4}D_{\frac{7}{2}}$ & $\frac{7}{2}^{+}$ &  7798 & 7687 & 7784 & 7724 &&&&\\
				\hline
				$3^{2}D_{\frac{3}{2}}$ & $\frac{3}{2}^{+}$ & 8049 &&&&&&&\\
				$3^{2}D_{\frac{5}{2}}$ & $\frac{5}{2}^{+}$ & 8057 &&&&&&&\\
				$3^{4}D_{\frac{1}{2}}$ & $\frac{1}{2}^{+}$ & 8041 &&&&&&&\\
				$3^{4}D_{\frac{3}{2}}$ & $\frac{3}{2}^{+}$ & 8047 &&&&&&&\\
				$3^{4}D_{\frac{5}{2}}$ & $\frac{5}{2}^{+}$ & 8055 &&&&&&&\\
				$3^{4}D_{\frac{7}{2}}$ & $\frac{7}{2}^{+}$ & 8063 &&&&&&&\\
				\hline
				\\
				\\
				\\
				$4^{4}D_{\frac{3}{2}}$ & $\frac{3}{2}^{+}$ & 8301 &&&&&&&\\
				$4^{2}D_{\frac{5}{2}}$ & $\frac{5}{2}^{+}$ & 8308 &&&&&&&\\
				$4^{4}D_{\frac{1}{2}}$ & $\frac{1}{2}^{+}$ & 8293 &&&&&&&\\
				$4^{4}D_{\frac{3}{2}}$ & $\frac{3}{2}^{+}$ & 8299 &&&&&&&\\
				$4^{4}D_{\frac{5}{2}}$ & $\frac{5}{2}^{+}$ & 8306 &&&&&&&\\
				$4^{4}D_{\frac{7}{2}}$ & $\frac{7}{2}^{+}$ & 8313 &&&&&&&\\
				\hline
				$5^{2}D_{\frac{3}{2}}$ & $\frac{3}{2}^{+}$ & 8541 &&&&&&&\\
				$5^{2}D_{\frac{5}{2}}$ & $\frac{5}{2}^{+}$ & 8548 &&&&&&&\\
				$5^{4}D_{\frac{1}{2}}$ & $\frac{1}{2}^{+}$ & 8534 &&&&&&&\\
				$5^{4}D_{\frac{3}{2}}$ & $\frac{3}{2}^{+}$ & 8539 &&&&&&&\\
				$5^{4}D_{\frac{5}{2}}$ & $\frac{5}{2}^{+}$ & 8546 &&&&&&&\\
				$5^{4}D_{\frac{7}{2}}$ & $\frac{7}{2}^{+}$ & 8553 &&&&&&&\\
				\hline
				\hline
				$1^{2}F_{\frac{5}{2}}$ & $\frac{5}{2}^{-}$ & 7675 & 7533 & 7702 & 7744 &&&&\\
				$1^{2}F_{\frac{7}{2}}$ & $\frac{7}{2}^{-}$ & 7685 & 7526 & 7685 & 7720  &&&&\\
				$1^{4}F_{\frac{3}{2}}$ & $\frac{3}{2}^{-}$ & 7663 & 7541 & 7721 & 7792 &&&&\\
				$1^{4}F_{\frac{5}{2}}$ & $\frac{5}{2}^{-}$ & 7673 & 7535 & 7708 & 7768 &&&&\\
				$1^{4}F_{\frac{7}{2}}$ & $\frac{7}{2}^{-}$ & 7683 & 7528 & 7690 & 7723 &&&&\\
				$1^{4}F_{\frac{9}{2}}$ & $\frac{9}{2}^{-}$ & 7694 & 7519 & 7670 & 7681 &&&&\\
				\hline
				$2^{2}F_{\frac{5}{2}}$ & $\frac{5}{2}^{-}$ & 7946 & 7758 & 7949 & 7956 &&&&\\
				$2^{2}F_{\frac{7}{2}}$ & $\frac{7}{2}^{-}$ & 7955 & 7753 & 7938 & 7931 &&&&\\
				$2^{4}F_{\frac{3}{2}}$ & $\frac{3}{2}^{-}$ & 7935 & 7764 & 7965 & 8001 &&&&\\
				$2^{4}F_{\frac{5}{2}}$ & $\frac{5}{2}^{-}$ & 7943 & 7760 & 7953 & 7978 &&&&\\
				$2^{4}F_{\frac{7}{2}}$ & $\frac{7}{2}^{-}$ & 7953 & 7755 & 7934 & 7935 &&&&\\
				$2^{4}F_{\frac{9}{2}}$ & $\frac{9}{2}^{-}$ & 7963 & 7749 & 7921 & 7892 &&&&\\
				\midrule
				\bottomrule
	\end{longtable}
\noindent The $\Omega_{bc}$ baryon is a doubly heavy baryon comprising a bottom, charm, and strange quark (bcs) in its quark configuration. Owing to this combination, it is electrically neutral and possesses an isospin of I=0. \\
\noindent In the present study, we have computed the mass spectra of the $\Omega_{bc}$ baryon across a broad range of states. The results of these calculations are detailed in Table \ref{tab2}. Our predicted ground state is consistent with various theoretical studies, showing close agreement with the results reported in Ref. \cite{Yoshida:2015tia} and exhibiting a small deviation of approximately 10 MeV from Ref. \cite{Roberts:2007ni}. The first orbital excitation, with spin-parity quantum numbers $J^{P}=\frac{1}{2}^{-}$ and $J^{P}=\frac{3}{2}^{-}$, demonstrate strong concordance with previous predictions, differing by only 9 MeV and 3 MeV from \cite{kakadiya2023spectroscopic} and \cite{Yoshida:2015tia}, respectively. The second excited states, corresponding to $J^{P}=\frac{3}{2}^{+}$ and $J^{P}=\frac{5}{2}^{+}$, also show good agreement with existing theoretical models, with deviations of 45 MeV and 23 MeV compared to \cite{kakadiya2023spectroscopic}. Additionally, the third orbital excitation, characterized by $J^{P}=\frac{5}{2}^{-}$ and $J^{P}=\frac{7}{2}^{-}$, are in reasonable alignment with previous predictions, differing by 27 MeV and 23 MeV from \cite{kakadiya2023spectroscopic} and \cite{Salehi:2019kvb}, respectively.

\subsection{$\Omega_{bb}$ Baryon}

\begin{longtable}{cccccccccccc}
	\caption{Predicted masses for $\Omega_{bb}$ states of ours and other approaches (in MeV); Here, \cite{kakadiya2023spectroscopic}, \cite{Shah:2016vmd} are our hCQM based with different potentials; \cite{Oudichhya:2021yln} is Regge Phenomenology; \cite{Ortiz-Pacheco:2023kjn} is non-relativistic harmonic oscillator model; \cite{Salehi:2019kvb}, \cite{Yoshida:2015tia}, \cite{Ghalenovi:2024scv}, \cite{Roberts:2007ni} are various constituent quark based model}
	\label{tab3}\\
	\toprule
	State &  $J^{P}$ & $Mass_{theo.}$ & \cite{kakadiya2023spectroscopic} & \cite{Shah:2016vmd} & \cite{Salehi:2019kvb} & \cite{Yoshida:2015tia} & \cite{Roberts:2007ni} & \cite{Ortiz-Pacheco:2023kjn} & \cite{Ghalenovi:2024scv} & \cite{Oudichhya:2021yln} & \cite{Tang:2011fv}\\
	\midrule
	\endfirsthead
	\caption[]{Predicted masses for $\Omega_{bb}$ states of ours and other approaches (in MeV); Here, \cite{kakadiya2023spectroscopic}, \cite{Shah:2016vmd} are our hCQM based with different potentials; \cite{Oudichhya:2021yln} is Regge Phenomenology; \cite{Ortiz-Pacheco:2023kjn} is non-relativistic harmonic oscillator model; \cite{Salehi:2019kvb}, \cite{Yoshida:2015tia}, \cite{Ghalenovi:2024scv}, \cite{Roberts:2007ni} are various constituent quark based model (continued)}\\
	\toprule
	State &  $J^{P}$ & $Mass_{theo.}$ & \cite{kakadiya2023spectroscopic} & \cite{Shah:2016vmd} & \cite{Salehi:2019kvb} & \cite{Yoshida:2015tia} & \cite{Roberts:2007ni} & \cite{Ortiz-Pacheco:2023kjn} & \cite{Ghalenovi:2024scv} & \cite{Oudichhya:2021yln} & \cite{Tang:2011fv}\\
	\midrule
	\endhead
$1^{2}S_{\frac{1}{2}}$ & $\frac{1}{2}^{+}$ & 10354 & 10357 & 10346 & 10870 & 10447 & 10544 & 10438 & 10400 & 10350 & 9890\\
$1^{4}S_{\frac{3}{2}}$ & $\frac{3}{2}^{+}$ & 10375 & 10406 & 10410 & 10866 & 10467 & 10486 & 10460 & & 10449 & 9930\\
\hline
$2^{2}S_{\frac{1}{2}}$ & $\frac{1}{2}^{+}$ & 10735 & 10646 & 10730 & 10969 & 10707 & 10693 &&& \\
			$2^{4}S_{\frac{3}{2}}$ & $\frac{3}{2}^{+}$ & 10749 & 10662 & 10737 & 10964 & 10723 & 10721 &&& \\
			\hline
			$3^{2}S_{\frac{1}{2}}$ & $\frac{1}{2}^{+}$ & 11052 & 10859 & 10973 & 11036 & 10744 & & & & 11109 \\
			$3^{4}S_{\frac{3}{2}}$ & $\frac{3}{2}^{+}$ & 11062 & 10866 & 10976 & 11032 & 10730 & & & & 11029 \\
			\hline
			$4^{2}S_{\frac{1}{2}}$ & $\frac{1}{2}^{+}$ & 11352 & 11039 & 11191 & 11123 & 10994 & & & &  \\
			$4^{4}S_{\frac{3}{2}}$ & $\frac{3}{2}^{+}$ & 11360 & 11043 & 11193 & 11119 & 11031 &&&& \\
			\hline
			$5^{2}S_{\frac{1}{2}}$ & $\frac{1}{2}^{+}$ & 11641 & 11196 & 11393 & 11230 &&&&& \\
			$5^{4}S_{\frac{3}{2}}$ & $\frac{3}{2}^{+}$ & 11647 & 11199 & 11394 & 11225 &&&&& \\
			\hline
			\hline
			$1^{2}P_{\frac{1}{2}}$ & $\frac{1}{2}^{-}$ & 10631 & 10580 & 10634 & 10968 & 10607 & 10616 & 10554 &&\\
			$1^{2}P_{\frac{3}{2}}$ & $\frac{3}{2}^{-}$ & 10638 & 10578 & 10629 & 10957 & 10608 & 10619 & 10560 & & 10638 \\
			$1^{4}P_{\frac{1}{2}}$ & $\frac{1}{2}^{-}$ & 10629 & 10581 & 10636 & 10976 & & & 10778 && \\
			$1^{4}P_{\frac{3}{2}}$ & $\frac{3}{2}^{-}$ & 10638 & 10579 & 10631 & 10963 & &  & 10784 && \\
			$1^{4}P_{\frac{5}{2}}$ & $\frac{5}{2}^{-}$ & 10642 & 10576 & 10625 & 10956 & & & 10795 & & 10729\\
			\hline
			$2^{2}P_{\frac{1}{2}}$ & $\frac{1}{2}^{-}$ & 10949 & 10796 & 10881 & 11031 & 10796 &&&& \\
			$2^{2}P_{\frac{3}{2}}$ & $\frac{3}{2}^{-}$ & 10955 & 10795 & 10881 & 11031 & 10797 &&&& \\
			$2^{4}P_{\frac{1}{2}}$ & $\frac{1}{2}^{-}$ & 10947 & 10797 & 10883 & 11039 &&&& &\\
			$2^{4}P_{\frac{3}{2}}$ & $\frac{3}{2}^{-}$ & 10954 & 10796 & 10879 & 11035 &&&&& \\
			$2^{4}P_{\frac{5}{2}}$ & $\frac{5}{2}^{-}$ & 10959 & 10794 & 10874 & 11019 &&&&& \\
			\hline
			\\
			\\
			$3^{2}P_{\frac{1}{2}}$ & $\frac{1}{2}^{-}$ &  11251 & 10982 & 11104 & 11124 & & & & & 11142 \\
			$3^{2}P_{\frac{3}{2}}$ & $\frac{3}{2}^{-}$ &  11256 & 10981 & 11101 & 11111 &&&&& \\
			$3^{4}P_{\frac{1}{2}}$ & $\frac{1}{2}^{-}$ &  11249 & 10982 & 11106 & 11131 &&&&& \\
			$3^{4}P_{\frac{3}{2}}$ & $\frac{3}{2}^{-}$ &  11255 & 10981 & 11103 & 11119 &&&&& \\
			$3^{4}P_{\frac{5}{2}}$ & $\frac{5}{2}^{-}$ &  11259 & 10980 & 11098 & 11112 &&&&& \\
			\hline
			$4^{2}P_{\frac{1}{2}}$ & $\frac{1}{2}^{-}$ & 11541 &&&&&&&&\\
			$4^{2}P_{\frac{3}{2}}$ & $\frac{3}{2}^{-}$ & 11545 &&&&&&&&\\
			$4^{4}P_{\frac{1}{2}}$ & $\frac{1}{2}^{-}$ & 11539 &&&&&&&&\\
			$4^{4}P_{\frac{3}{2}}$ & $\frac{3}{2}^{-}$ &  11544  &&&&&&&&\\
			$4^{4}P_{\frac{5}{2}}$ & $\frac{5}{2}^{-}$ &  11549 &&&&&&&&\\
			\hline
			$5^{2}P_{\frac{1}{2}}$ & $\frac{1}{2}^{-}$ & 11821 &&&&&&&&\\
			$5^{2}P_{\frac{3}{2}}$ & $\frac{3}{2}^{-}$ & 11825 &&&&&&&&\\ 
			$5^{4}P_{\frac{1}{2}}$ & $\frac{1}{2}^{-}$ &  11819 &&&&&&&&\\
			$5^{4}P_{\frac{3}{2}}$ & $\frac{3}{2}^{-}$ & 11824 &&&&&&&&\\
			$5^{4}P_{\frac{5}{2}}$ & $\frac{5}{2}^{-}$ &  11829 &&&&&&&&\\
			\hline
			\hline
			$1^{2}D_{\frac{3}{2}}$ & $\frac{3}{2}^{+}$ &  10843 & 10660 & 10783 & 11032 &&&&&\\
			$1^{2}D_{\frac{5}{2}}$ & $\frac{5}{2}^{+}$ &  10849 & 10657 & 10777 & 11017 & & & & & 10918 \\
			$1^{4}D_{\frac{1}{2}}$ & $\frac{1}{2}^{+}$ &  10835 & 10662 & 10789 & 11017 &&&&&\\
			$1^{4}D_{\frac{3}{2}}$ & $\frac{3}{2}^{+}$ &  10841 & 10661 & 10785 & 11033 &&&&&\\
			$1^{4}D_{\frac{5}{2}}$ & $\frac{5}{2}^{+}$ &  10848 & 10658 & 10779 & 11019 &&&&&\\
			$1^{4}D_{\frac{7}{2}}$ & $\frac{7}{2}^{+}$ &  10854 & 10655 & 10772 & 11007 & & & & & 11002 \\
			\hline
			$2^{2}D_{\frac{3}{2}}$ & $\frac{3}{2}^{+}$ &  11147 & 10864 & 11012 & 11116 &&&&&\\
			$2^{2}D_{\frac{5}{2}}$ & $\frac{5}{2}^{+}$ &  11153 & 10863 & 11008 & 11104 & & & & & 11025 \\
			$2^{4}D_{\frac{1}{2}}$ & $\frac{1}{2}^{+}$ &  11140 & 10866 & 11017 & 11126 &&&&&\\
			$2^{4}D_{\frac{3}{2}}$ & $\frac{3}{2}^{+}$ &  11145 & 10865 & 11014 & 11119 &&&&&\\
			$2^{4}D_{\frac{5}{2}}$ & $\frac{5}{2}^{+}$ &  11152 & 10863 & 11009 & 11106 &&&&&\\
			$2^{4}D_{\frac{7}{2}}$ & $\frac{7}{2}^{+}$ &  11157 & 10861 & 11004 & 11094 & & & & & 11021 \\
			\hline
			\\
			\\
			\\
			\\
			$3^{2}D_{\frac{3}{2}}$ & $\frac{3}{2}^{+}$ & 11439 &&&&&&&&\\
			$3^{2}D_{\frac{5}{2}}$ & $\frac{5}{2}^{+}$ & 11445 &&&&&&&&\\
			$3^{4}D_{\frac{1}{2}}$ & $\frac{1}{2}^{+}$ & 11433 &&&&&&&&\\
			$3^{4}D_{\frac{3}{2}}$ & $\frac{3}{2}^{+}$ & 11438 &&&&&&&&\\
			$3^{4}D_{\frac{5}{2}}$ & $\frac{5}{2}^{+}$ & 11443 &&&&&&&&\\
			$3^{4}D_{\frac{7}{2}}$ & $\frac{7}{2}^{+}$ & 11449 &&&&&&&&\\
			\hline
			$4^{4}D_{\frac{3}{2}}$ & $\frac{3}{2}^{+}$ & 11722 &&&&&&&&\\
			$4^{2}D_{\frac{5}{2}}$ & $\frac{5}{2}^{+}$ & 11727 &&&&&&&&\\
			$4^{4}D_{\frac{1}{2}}$ & $\frac{1}{2}^{+}$ & 11716 &&&&&&&&\\
			$4^{4}D_{\frac{3}{2}}$ & $\frac{3}{2}^{+}$ & 11720 &&&&&&&&\\
			$4^{4}D_{\frac{5}{2}}$ & $\frac{5}{2}^{+}$ & 11725 &&&&&&&&\\
			$4^{4}D_{\frac{7}{2}}$ & $\frac{7}{2}^{+}$ & 11731 &&&&&&&&\\
			\hline
			$5^{2}D_{\frac{3}{2}}$ & $\frac{3}{2}^{+}$ & 11996 &&&&&&&&\\
			$5^{2}D_{\frac{5}{2}}$ & $\frac{5}{2}^{+}$ & 12001 &&&&&&&&\\
			$5^{4}D_{\frac{1}{2}}$ & $\frac{1}{2}^{+}$ & 11991 &&&&&&&&\\
			$5^{4}D_{\frac{3}{2}}$ & $\frac{3}{2}^{+}$ & 11995 &&&&&&&&\\
			$5^{4}D_{\frac{5}{2}}$ & $\frac{5}{2}^{+}$ & 12000 &&&&&&&&\\
			$5^{4}D_{\frac{7}{2}}$ & $\frac{7}{2}^{+}$ & 12005 &&&&&&&&\\
			\hline
			\hline
			$1^{2}F_{\frac{5}{2}}$ & $\frac{5}{2}^{-}$ & 11042 & 10727 & 10920 & 11105 &&&&&\\
			$1^{2}F_{\frac{7}{2}}$ & $\frac{7}{2}^{-}$ & 11049 & 10724 & 10913 & 11082 & & & & & 11191\\
			$1^{4}F_{\frac{3}{2}}$ & $\frac{3}{2}^{-}$ & 11033 & 10730 & 10927 & 11118 &&&&&\\
			$1^{4}F_{\frac{5}{2}}$ & $\frac{5}{2}^{-}$ & 11040 & 10728 & 10922 & 11107 &&&&&\\
			$1^{4}F_{\frac{7}{2}}$ & $\frac{7}{2}^{-}$ & 11047 & 10725 & 10915 & 11094 &&&&&\\
			$1^{4}F_{\frac{9}{2}}$ & $\frac{9}{2}^{-}$ & 11055 & 10721 & 10907 & 11076 &&&&&\\
			\hline
			$2^{2}F_{\frac{5}{2}}$ & $\frac{5}{2}^{-}$ & 11336 & 10923 & 11136 & 11210 &&&&&\\
			$2^{2}F_{\frac{7}{2}}$ & $\frac{7}{2}^{-}$ & 11343 & 10921 & 11130 & 11189 &&&&&\\
			$2^{4}F_{\frac{3}{2}}$ & $\frac{3}{2}^{-}$ & 11329 & 10925 & 11142 & 11226 &&&&&\\
			$2^{4}F_{\frac{5}{2}}$ & $\frac{5}{2}^{-}$ & 11335 & 10923 & 11137 & 11214 &&&&&\\
			$2^{4}F_{\frac{7}{2}}$ & $\frac{7}{2}^{-}$ & 11341 & 10921 & 11132 & 11202 &&&&&\\
			$2^{4}F_{\frac{9}{2}}$ & $\frac{9}{2}^{-}$ & 11349 & 10919 & 11125 & 11179 & & & & & 11268 \\
			\midrule
			\bottomrule
	\end{longtable}

\noindent The $\Omega_{bb}$ baryon is a doubly heavy baryon composed of two bottom quarks and one strange quark (bbs). Due to its quark content, it carries a negative electric charge and has an isospin of I=0. In this study, we have investigated the mass spectra of the $\Omega_{bb}$ baryon across a comprehensive range of excited states. The detailed results of these calculations are provided in table \ref{tab3}.\\

\noindent As with other members of the doubly heavy $\Omega$ baryon family, no experimental evidence has yet been reported for the $\Omega_{bb}$baryon or its excited states. Nonetheless, our predicted ground state exhibits strong agreement with existing theoretical models, closely matching the results in Ref. \cite{kakadiya2023spectroscopic} and differing by approximately 8 MeV from those in Ref. \cite{Shah:2016vmd}.
The first orbital excitation, with spin-parity quantum numbers $J^{P}=\frac{1}{2}^{-}$ and $J^{P}=\frac{3}{2}^{-}$, are found to be in close correspondence with previous predictions, showing differences of 3 MeV and 24 MeV when compared to Refs. \cite{kakadiya2023spectroscopic} and \cite{Yoshida:2015tia}, respectively. Similarly, the second orbital excitation with $J^{P}=\frac{3}{2}^{+}$ and $J^{P}=\frac{5}{2}^{+}$ display deviations of 63 MeV and 42 MeV from Ref. \cite{Shah:2016vmd}. Furthermore, the third orbital excitation, characterized by $J^{P}=\frac{5}{2}^{-}$ and $J^{P}=\frac{7}{2}^{-}$, show differences of 63 MeV and 41 MeV relative to Refs. \cite{kakadiya2023spectroscopic} and \cite{Salehi:2019kvb}, respectively. These results not only provide a detailed theoretical framework for the mass spectrum of the $\Omega_{bb}$ baryon but also serve as valuable benchmarks for future experimental investigations in the field of heavy baryon spectroscopy.\\

As highlighted in the methodology section, there are parameters on which the obtained resonance masses depend. The present manuscript is an update on our earlier works where in the mass hierarchy was not obtained. The incorporation of second-order mass correction in the L-S coupling parts have resolved this. So, the constituent quark masses have not been changed here so as to keep in accordance with our previous works. 
The fitting parameters $\upsilon$ and $\sigma_{1}$ have been varied with nearly $\pm5\%$ which changes the value by a few MeV. However, this part is incorporated from pNRQCD model, so parameters are kept consistent. But this factor will not be affecting in the ground state or S-wave due to implementation in only L-S coupling part.

Thus, the string tension of the confinement potential has been varied in the range of $\pm5\%$ to $\pm10\%$. The confinement constant is nearly 0.12 for the present case of $\Omega_{cc}$ which we have checked for 0.09, 0.12 and 0.2 and the results are shown in the table.\\
\begin{table}[h]
	\caption{Uncertainty Analysis of fitting parameters}
	\begin{tabular}{|c|c|c|c|}
		\hline
		State & \multicolumn{3}{c|}{Confinement term values} \\
		\hline
		 & 0.09 & 0.12 & 0.2 \\
		 \hline
		1S & 3569  &  3692  &  3940 \\
		1S & 3631 & 3780 & 4048 \\
		\hline
		2S & 3871 & 4094 & 4464 \\
		2S & 3911 & 4148 & 4544 \\
		\hline
		3S & 4135 & 4427 & 4913 \\
		3S & 4162 & 4464 & 4967 \\
		\hline
		
		\end{tabular}
\end{table}

As the ground state itself is not known experimentally, we have taken up the lattice QCD masses as reference to obtain the parameters so as to reproduce it. As it is evident from the table, a lower side of the parameter, under-predicts the ground state which ultimately travels through all the higher states. As well as the higher value of this parameter, over-predicts the ground state. Also, it is noteworthy that the splitting between these spin-states increases with increase in the confinement parameter. This shall diverge the masses at even higher orbital states. Similar trend is followed in all obtained $\Omega_{QQ^{'}}$ masses. 

The current resonance masses presented are also in accordance with those of Regge phenomenology which is also aiding us to compare the ground state. The uncertainty analysis would be more precise when we have more experimentally established states to compare our masses. As this is a phenomenology study, the experimental values and more predicted masses from other models will lead towards the improvement of the present study.

\section{Regge Trajectories}\label{sec4}
\begin{figure}[hbt!]
	\centering
	\fbox{\includegraphics[scale=0.25]{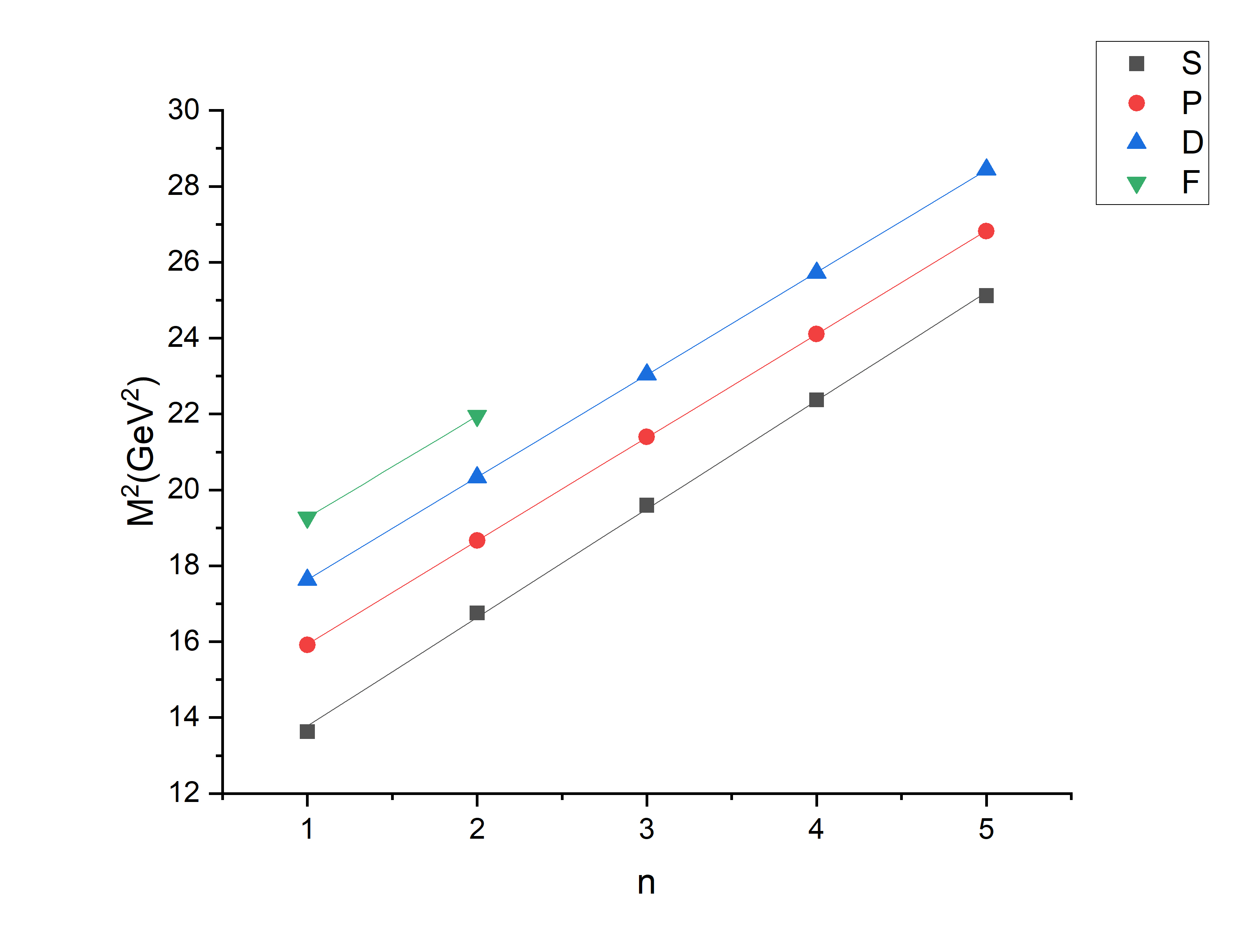}}
	\centering
	\caption{Plot of principle quantum number $n$ vs $M^{2}$ based on the predicted masses of  $\Omega_{cc}$ baryon for different orbitals}\label{fig1}
\end{figure}
\begin{figure}[hbt!]
	\centering
	\fbox{\includegraphics[scale=0.25]{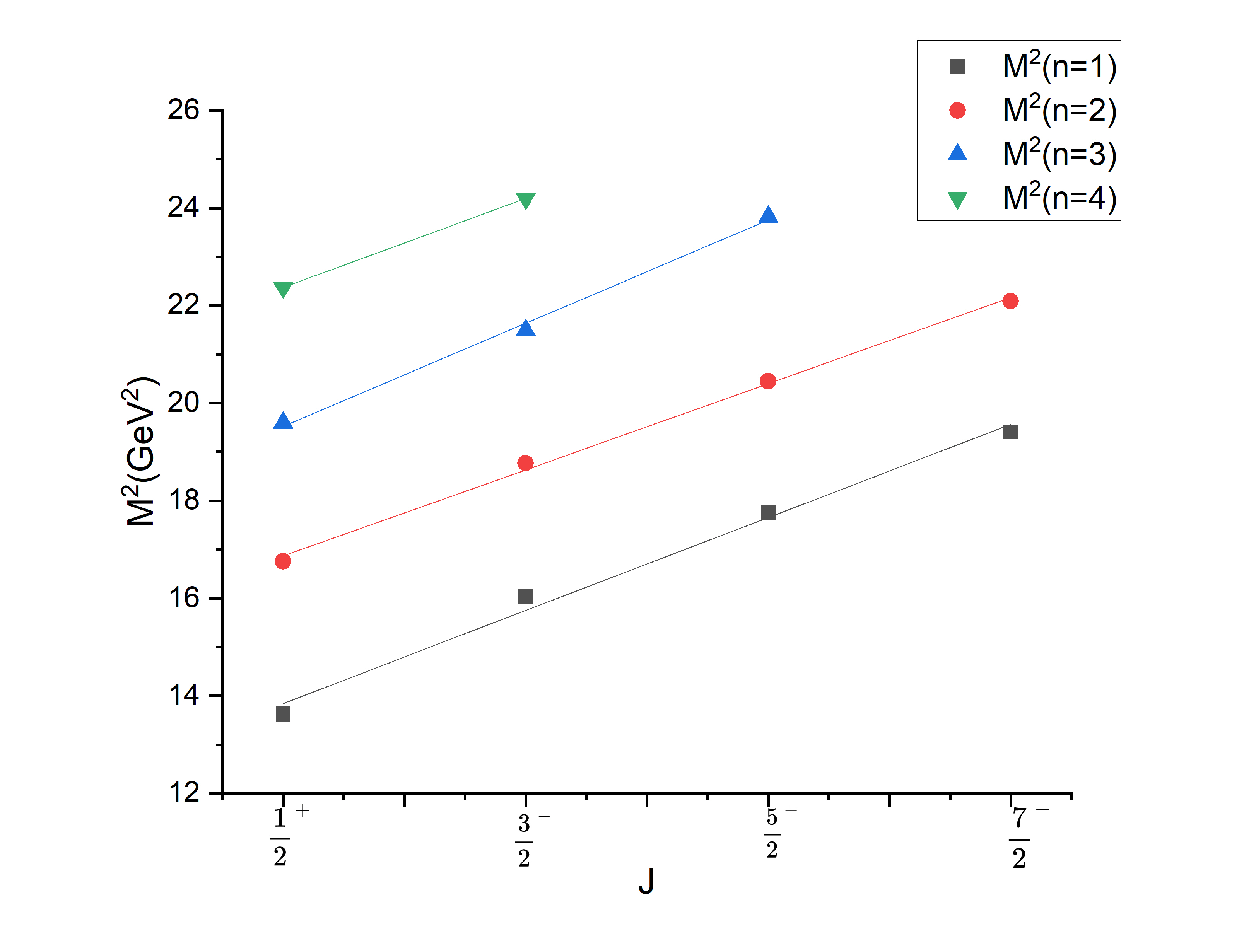}}
	\centering
	\caption{Plot of spin parity $J^P$ vs $M^{2}$ of $\Omega_{cc}$ for spin $\frac{1}{2}$}\label{fig2}
\end{figure}
\begin{figure}[hbt!]
	\centering
	\fbox{\includegraphics[scale=0.25]{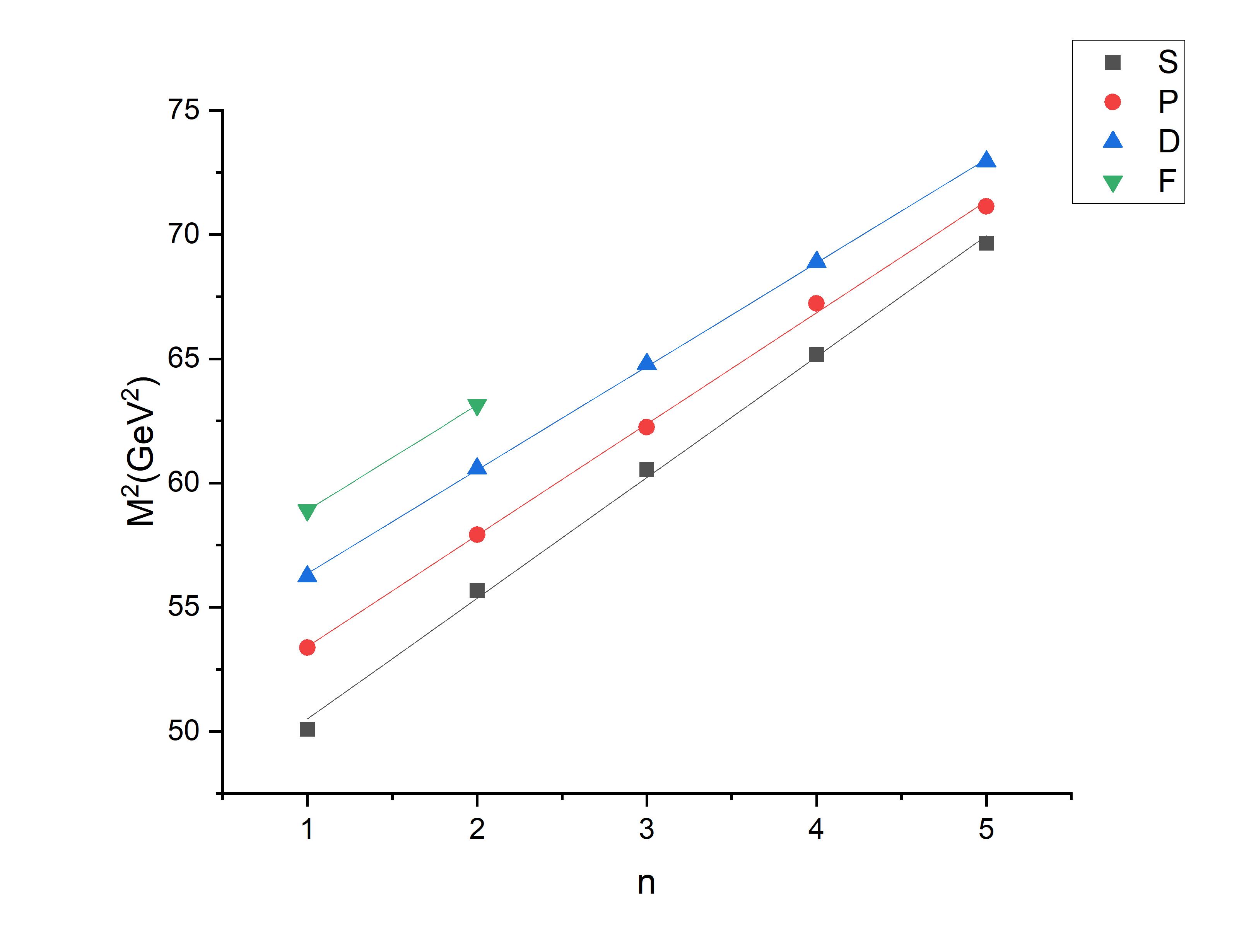}}
	\caption{Plot of principle quantum number $n$ vs $M^{2}$ based on the predicted masses of  $\Omega_{bc}$ baryon for different orbitals}\label{fig3}
\end{figure}

One of the helpful tools in spectroscopic research has been Regge trajectories. Chew and Frautschi applied the Regge poles theory to investigate the analyticity
in the case of strong interactions. They postulated that all strongly interacting particles are self-generating (the bootstrap hypothesis) and that they must lie on Regge trajectories (Chew-Frautschi conjecture).
Regge trajectories provide a powerful graphical representation in hadron spectroscopy, illustrating the relationship between a particle's total angular momentum (denoted by J) or its radial quantum number (denoted by n) and the square of its mass ($M^{2}$). These plots typically J vs $M^{2}$) and (n vs $M^{2}$) exhibit a striking linearity, where each trajectory appears as a straight line. Moreover, these lines are found to be approximately parallel and equidistant from one another, indicating a regular and predictable pattern in the excitation spectrum of hadrons.

\begin{equation}
	n=c M^{2} + c_{0}
\end{equation}
\begin{equation}
	J=d M^{2} + d_{0}
\end{equation}

\noindent Here, n=1,2,3 represents the radial principal quantum number, and J denotes the total angular momentum. The trajectories are characterized by c, d, $c_{0}$ and $d_{0}$, where c and d are the slope and $c_{0}$ and $d_{0}$ are intercept, respectively, of the trajectories.\\
To validate our calculations, Regge trajectory based on our predicted masses has been plotted to check for the linear behavior which shall aid in assigning the spin-parity of yet unknown states. 
\begin{figure}[hbt!]
	\centering
	\fbox{\includegraphics[scale=0.25]{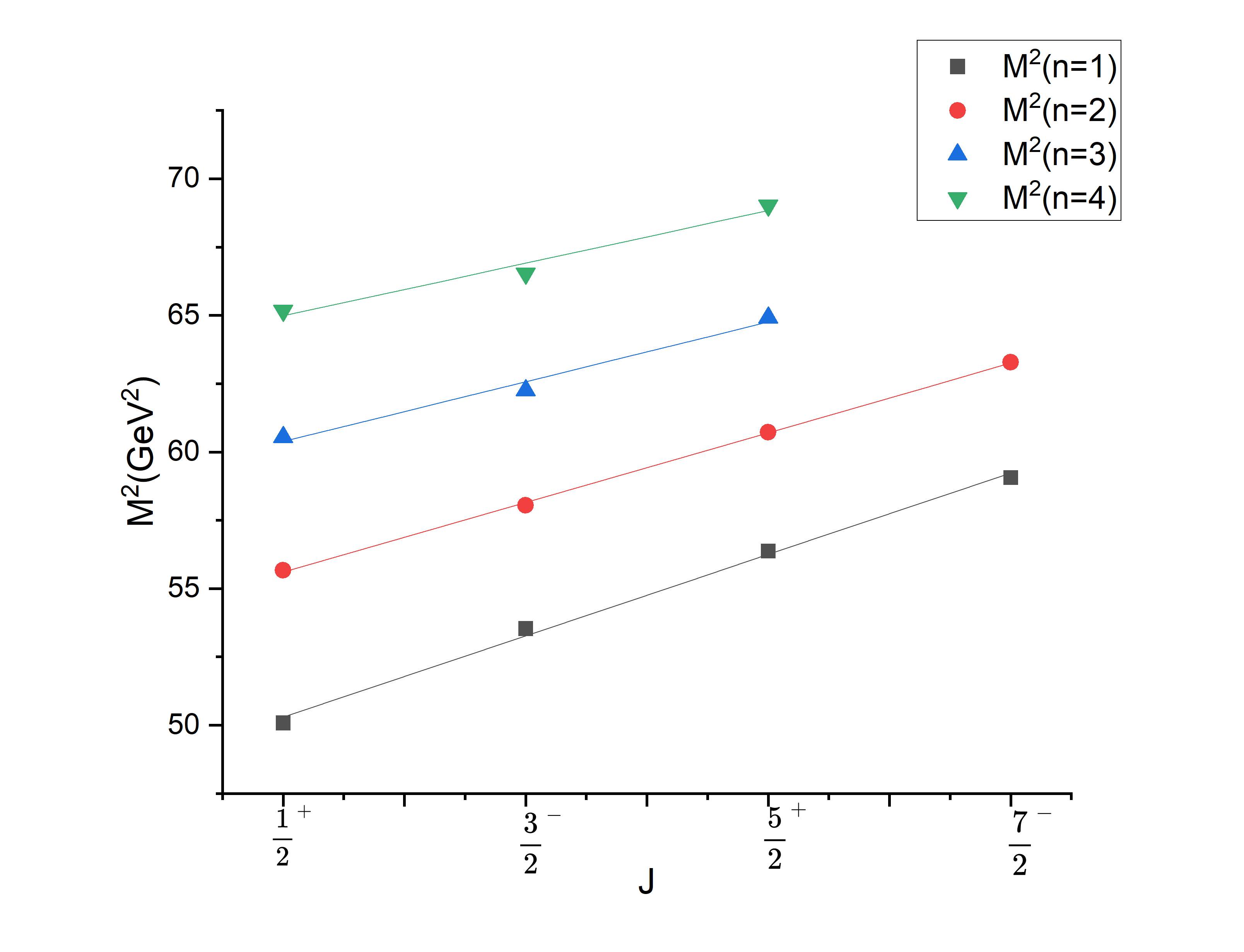}}
	\caption{Plot of spin parity $J^P$ vs $M^{2}$ of $\Omega_{bc}$ for spin $\frac{1}{2}$} \label{fig4}
\end{figure}
\begin{figure}[hbt!]
	\centering
	\fbox{\includegraphics[scale=0.25]{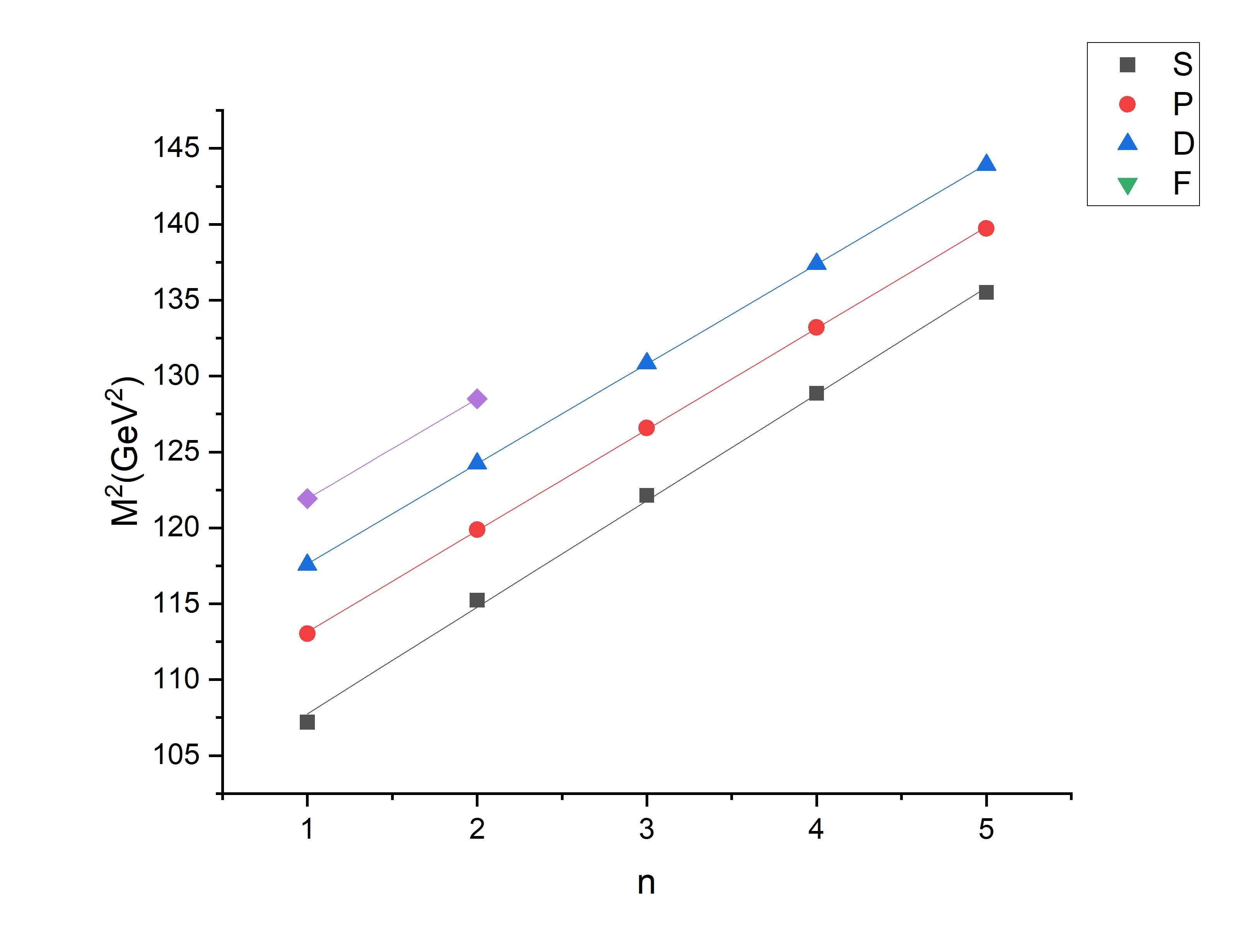}}
	\caption{Plot of principle quantum number $n$ vs $M^{2}$ based on the predicted masses of  $\Omega_{bc}$ baryon for different orbitals}\label{fig5}
\end{figure}

\begin{figure}[hbt!]
	\centering
	\fbox{\includegraphics[scale=0.25]{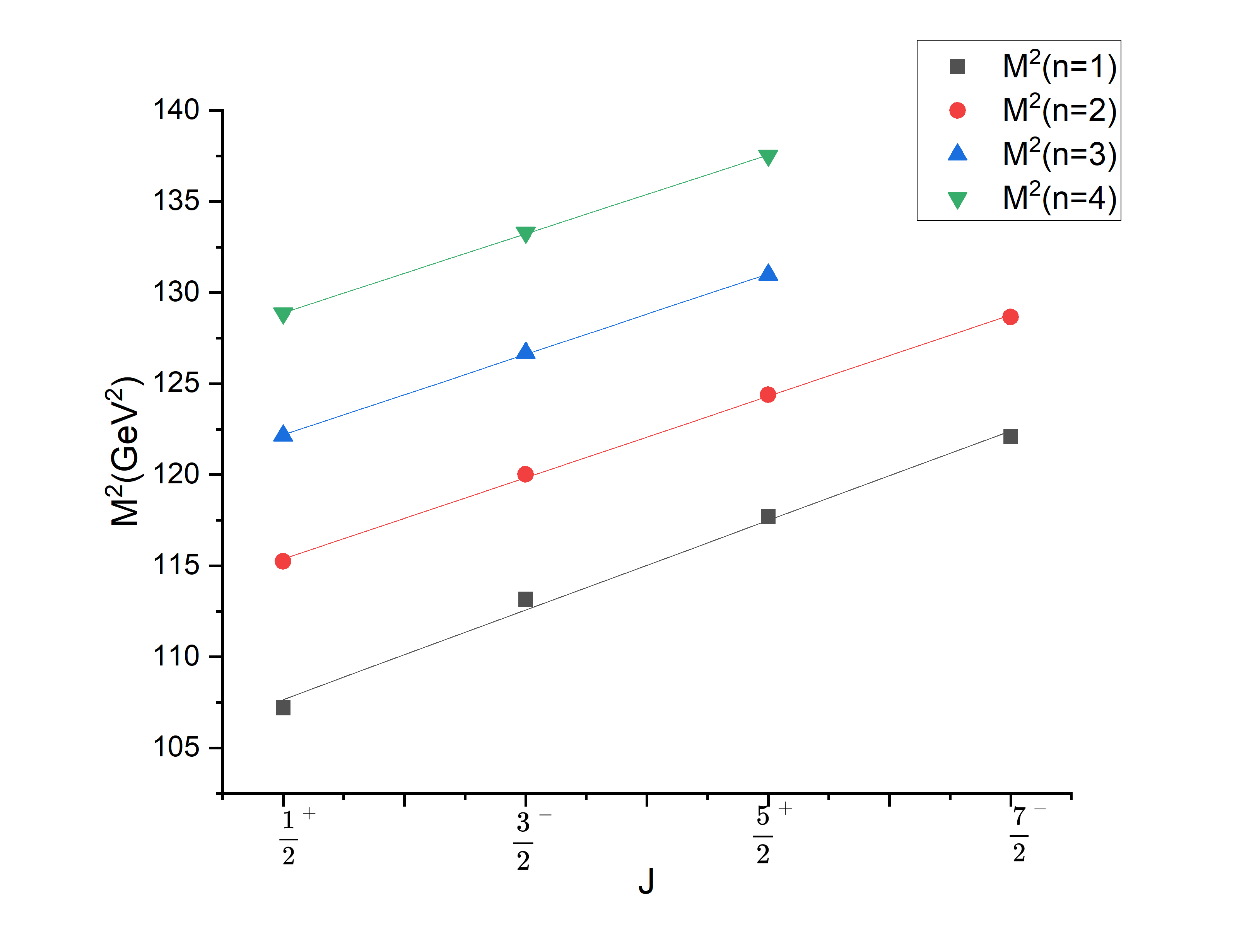}}
	\caption{Plot of spin parity $J^P$ vs $M^{2}$ of $\Omega_{bb}$ for spin $\frac{1}{2}$} \label{fig6}
\end{figure}
\newpage
\section{Magnetic Moment}\label{sec5}
The magnetic moment plays a vital role in understanding the internal dynamics and quark composition of baryons. Numerous theoretical approaches have been developed to analyze the magnetic moments of such doubly heavy baryons \cite{Ozdem:2019zis,Ozdem:2023cri}. In general, the magnetic moment of a baryon is determined by key factors: the electric charge of the constituent quarks, their effective masses, and the spin-flavor structure described by the baryon's wave function. A commonly used formula to calculate the magnetic moment combines these elements, allowing for reliable theoretical predictions that can be compared with experimental data.
\begin{equation}
	{\mu_{B}}=\sum_{j} <\phi_{sf}|\hat{\mu_{qz}}|\phi_{sf}>
\end{equation}
Here,  $\phi_{sf}$ signifies the spin-flavor wave function of the baryon, and $\hat{\mu_{qz}}$ is indicative of the magnetic moment operator. 
\begin{longtable}{cccccccccc}
	\caption{Predicted magnetic moment(in nuclear magneton) with spin flavour wave function of $j^{p}=\frac{1}{2}^{+}$ and $j^{p}=\frac{3}{2}^{+}$ doubly heavy baryons of ours and compared to other approaches}\label{tab4}\\
	\hline
	Baryons & function & Our & \cite{kakadiya2023spectroscopic} & \cite{Shah:2016vmd} & \cite{Patel:2007gx} & \cite{Albertus:2006ya} & \cite{Ozdem:2018uue} \\
	\hline
	\hline
	$\Omega_{cc}^{+}$ & $\frac{4}{3}$$\mu_{c}$$-$$\frac{1}{3}$$\mu_{s}$ & 0.712 & 0.704 & 0.692 & 0.785 &  0.635 & 0.39 $\pm$ 0.09\\
	$\Omega_{cb}^{0}$  & $\frac{2}{3}$$\mu_{c}$$+$$\frac{2}{3}$$\mu_{b}$$-$$\frac{1}{3}$$\mu_{s}$  & 0.446 & 0.447 & 0.439 & 0.397 & 0.368 \\
	$\Omega_{bb}^{-}$  & $\frac{4}{3}$$\mu_{b}$$-$$\frac{1}{3}$$\mu_{s}$ & 0.113 & 0.113 & 0.108 & 0.109 & 0.101 & 0.42 $\pm$ 0.05\\	
	$\Omega_{cc}^{*+}$  & 2$\mu_{c}$$+$$\mu_{s}$ & 0.286 & 0.283 & 0.285 & 0.121 & 0.139 \\	
	$\Omega_{cb}^{*0}$  & $\mu_{c}$$+$$\mu_{s}$$+$$\mu_{b}$ & -0.182 & -0.181 & -0.181 & -0.217 & -0.261 \\
	$\Omega_{bb}^{*-}$  & 2$\mu_{b}$$+$$\mu_{s}$ & -0.718 & -0.718 & -0.839 & -0.711 & -0.662\\
	\hline
\end{longtable}

The specified reference provides the magnetic moment for an individual quark.
\begin{equation}
	\mu_{j}=\frac{e_{j}}{2m_{j}^{eff}}.\sigma_{j}
\end{equation}\\
Here, $e_{j}$ refers to the individual charge of a constituent quark, $\sigma_{j}$ represents the spin-flavor wave function, and $m_{j}^{eff}$ indicates the effective mass of the quark.

\noindent We computed the wave function \cite{universe7090337} and ground state of the magnetic moment in table \ref{tab4}, and the comparison with other theoretical models revealed variations within the range of 0.1 to 0.5 $\mu_{N}$, indicating an accurate prediction.\\

\section{Transition Magnetic Moment and Radiative Decay}\label{sec6}

In baryonic systems, the transition magnetic moment refers to the magnetic moment associated with radiative transitions between two different quantum states or energy levels of a baryon. Unlike the static magnetic moment, which pertains to a baryon in a specific state, the transition magnetic moment captures the magnetic properties involved in a quantum transition typically occurring during processes such as electromagnetic decays.\\
To compute the transition magnetic moment, one typically employs the spin-flavor wave functions of the initial ($B_{e}$) and final ($B_{e}^{'}$) baryon states. The calculation involves the expectation value of the z-component of the magnetic moment operator, taken between these two states.
\begin{equation}
	\mu_{Be}\longrightarrow\mu_{Be^{'}} = <\phi_{Be}|\mu_{Be}\rightarrow\mu_{Be^{'}}|\phi_{Be^{'}}>
\end{equation} 
Noteworthy that $\phi_{Be}$ represents the initial state of the spin-flavor wave function, while $\phi_{Be^{'}}$ denotes the final state of the spin-flavor wave function.\\

\noindent Radiative decay refers to the process wherein a particle transforms into another particle while simultaneously emitting a photon; this phenomenon is also known as electromagnetic decay.
The radiative decay width is expressed as follows: \cite{Gandhi:2018lez,Shah:2016nxi}
\begin{align}
	\Gamma_{\gamma} = \frac{k^3}{4\pi}\frac{2}{2J+1}\frac{e}{m^{2}_{p}}\mu^{2}_{B_{e}\rightarrow B^{'}_e},
\end{align}
In this context, k represents the energy of the photon, J corresponds to the total angular momentum, and $m_{p}$ denotes the mass of the proton. The results for the same are shown in table \ref{tab5}.\\
\newpage

\begin{longtable}{cccccccc}
	\caption{Radiative decay and the predicted transition magnetic moment (measured in nuclear magnetons) and their function.}\label{tab5}\\
	\hline
	Decay & Function & \multicolumn{2}{c}{Transition magnetic moment} & \multicolumn{2}{c}{ Radiative Decay} \\
	& & [our] & \cite{Hazra:2021lpa} & [our]& \cite{Li:2017pxa}\\
	\hline
	\hline
	
	$\Omega_{cc}^{*+}$$\longrightarrow$ $\Omega_{cc}^{+}$ $+$ $\gamma$ & $\frac{\sqrt{2}}{3}$($\mu_{s}$$-$$\mu_{b}$) & 0.912 & 0.96 &  2.247 & 1.973 \\
	$\Omega_{cb}^{*0}$$\longrightarrow$ $\Omega_{cb}^{0}$ $+$ $\gamma$ & $\frac{\sqrt{2}}{3}$($\mu_{s}$$-$$\mu_{b}$) & 0.723 & 0.69 & 1.12 & 0.89 \\
	$\Omega_{bb}^{*-}$$\longrightarrow$ $\Omega_{bb}^{-}$ $+$ $\gamma$ & $\frac{\sqrt{2}}{3}$($\mu_{s}$$-$$\mu_{b}$) & 0.492 & 0.48 & 0.10 & 0.09\\

	\hline	 
\end{longtable}

\section{Conclusion}\label{sec7}

In this study, we have systematically investigated the mass spectra of doubly heavy baryons within the framework of the Hypercentral Constituent Quark Model (hCQM). The model incorporates a Coulomb plus confining potential along with spin-orbit interactions and higher-order corrections up to $O(\frac{1}{m^{2}})$. We obtained precise predictions for both ground and excited (radial and orbital) states. The computed mass spectra show strong agreement with results from other established theoretical models.\\
In addition to the mass analysis, we calculated the magnetic moments of singly and doubly heavy baryons, taking into account the contributions from quark charges, effective masses, and spin-flavor wave functions. These magnetic moments provide deeper insight into the internal quark dynamics and static electromagnetic properties of heavy baryons.\\
We also explored transition magnetic moments, which are key to understanding radiative decays between different baryonic states. Using spin-flavor wave functions of the initial and final states, we determined the magnetic moment operators governing these transitions. These calculations are instrumental in predicting decay rates and selection rules for electromagnetic transitions, thus complementing our spectroscopic findings.\\
Furthermore, we assigned spin-parity quantum numbers ($J^{P}$) to both known and predicted states and validated the internal consistency of our model by plotting Regge trajectories in the (n,$M^{2}$) and (J,$M^{2}$) plane for the $\Omega_{cc}$, $\Omega_{bc}$, and $\Omega_{bb}$ baryons in figures [\ref{fig1},\ref{fig2},\ref{fig3},\ref{fig4},\ref{fig5},\ref{fig6}]. The observed linearity and regular spacing of these trajectories reinforce the predictive reliability of our approach. \\
\begin{itemize}
	\item As observed in the section \ref{sec3}, the present results incorporating the second order correction terms, updates the hierarchy of the assigned $J^P$ values for given radial state.
	\\
	\item  In case of $\Omega_{cc}$, the ground state 1S is in accordance with the lattice studies, also differs by nearly 50 MeV with harmonic oscillator model, Regge phenomenology. The same ground state is predicted to be 3630 MeV by QCD Sum Rule model \cite{Tang:2011fv} which is around 60 MeV apart. However, a notable point about the difference of $J^{P}=\frac{1}{2}^{+}$ and $J^{P}=\frac{3}{2}^{+}$ splitting is nearly same for our model as well as QCD Sum Rule and few other models. \\
	The first orbital excitation with spin-parity quantum numbers $J^{P}=\frac{1}{2}^{-}$ and $J^{P}=\frac{3}{2}^{-}$, which exhibit a high degree of consistency with prior theoretical predictions, differing by 21 MeV and 29 MeV from \cite{kakadiya2023spectroscopic} and \cite{Mathur:2002ce}, respectively. Similarly, the second orbital excitation with $J^{P}=\frac{3}{2}^{+}$ and $J^{P}=\frac{5}{2}^{+}$show deviations of 44 MeV and 65 MeV from \cite{Shah:2016vmd}, respectively. The third orbital states, with $J^{P}=\frac{5}{2}^{-}$ and $J^{P}=\frac{7}{2}^{-}$, also align well with the literature, exhibiting only minor differences of 1 MeV and 53 MeV when compared with \cite{Shah:2016vmd} and \cite{Salehi:2019kvb}. 
	\\
	\item In case of $\Omega_{bc}$, ground state itself varies with quite a large variations. However, our results are comparable to a few other CQMs but QCD Sum rules is under-predicted. The exact values of ground state as well as other excited states looks forward to new experimental results which would allow us to fit the model parameters for reproducing the known ground state. \\
	The first orbital excitation, with spin-parity quantum numbers $J^{P}=\frac{1}{2}^{-}$ and $J^{P}=\frac{3}{2}^{-}$, demonstrate strong concordance with previous predictions, differing by only 9 MeV and 3 MeV from \cite{kakadiya2023spectroscopic} and \cite{Yoshida:2015tia}, respectively. The second excited states, corresponding to $J^{P}=\frac{3}{2}^{+}$ and $J^{P}=\frac{5}{2}^{+}$, also show good agreement with existing theoretical models, with deviations of 45 MeV and 23 MeV compared to \cite{kakadiya2023spectroscopic}. \\
	Additionally, the third orbital excitation, characterized by $J^{P}=\frac{5}{2}^{-}$ and $J^{P}=\frac{7}{2}^{-}$, are in reasonable alignment with previous predictions, differing by 27 MeV and 23 MeV from \cite{kakadiya2023spectroscopic} and \cite{Salehi:2019kvb}, respectively. Other approaches have limited states available for the comparison.
	\\
	\item $\Omega_{bb}$ results show agreement in 1S state with a recent study using deep neural network as well as Regge phenomenology, but diverges significantly from a few approaches such as QCD Sum Rule. \\
	1P, 1D state varies within 100 MeV from few approaches. The first orbital excitation, with spin-parity quantum numbers $J^{P}=\frac{1}{2}^{-}$ and $J^{P}=\frac{3}{2}^{-}$, are found to be in close correspondence with previous predictions, showing differences of 3 MeV and 24 MeV when compared to Refs. \cite{kakadiya2023spectroscopic} and \cite{Yoshida:2015tia}, respectively. \\
	Similarly, the second orbital excitation with $J^{P}=\frac{3}{2}^{+}$ and $J^{P}=\frac{5}{2}^{+}$ display deviations of 63 MeV and 42 MeV from Ref. \cite{Shah:2016vmd}. Furthermore, the third orbital excitation, characterized by $J^{P}=\frac{5}{2}^{-}$ and $J^{P}=\frac{7}{2}^{-}$, show differences of 63 MeV and 41 MeV relative to Refs. \cite{kakadiya2023spectroscopic} and \cite{Salehi:2019kvb}, respectively.
	\\
	\item The broad comparison with various models have enabled us if not be with precise mass but a range for experimental findings too. The Regge trajectory shall allow us to place a given state on the linear paths corresponding to give the tentative spin-parity assignment. These comparisons shall help towards the possible modifications to be incorporated in models to reach the final spectra. Also, the hypercentral Constituent Quark Model itself accounts for all the effects inside a baryon, such as sea quark, etc. are all combined in constituent quark mass. But this study doesn't allow us to look for any anomalous behavior of a given state or precisely comment on the inside structure of quarks. 
\end{itemize}
Overall, this comprehensive analysis not only enhances the understanding of mass spectra but also sheds light on the electromagnetic structure and decay mechanisms of doubly heavy baryons, offering a robust theoretical foundation for guiding future experimental investigations. \\
The states predicted based on the current approach shall be validated with new experimental data. This model has allowed us to proceed phenomenologically deciding the parameters of the model with ground state and the expected difference between two spin states. The experimental results shall allow us to look for any other parameter to be added in our current understanding of this Constituent Quark Model. 

\subsubsection*{Funding Declaration}
Funding information - Not applicable 

\bibliography{sn-bibliography}

\end{document}